\documentclass{jpp}     

\usepackage[makeroom]{cancel}
\usepackage{scrextend} 
\usepackage{bold-extra} 
\usepackage{color}
\usepackage[T1]{fontenc} 
\usepackage{fancyhdr}
\usepackage{ulem}
\usepackage{xcolor}
\usepackage{relsize}
\usepackage{textcomp}
\usepackage{lipsum}
\usepackage{setspace}
\usepackage{empheq}

\usepackage{float} 
\usepackage{epstopdf, epsfig}
\usepackage{graphicx} 
\usepackage{graphics} 
\usepackage[justification=raggedright]{caption} 
\usepackage{tabu} 
\usepackage{tabulary} 
\usepackage{tabularx} 
\usepackage{booktabs}

\usepackage{framed} 
\usepackage{listings} 

\usepackage{amsmath} 
\usepackage{amssymb} 
\usepackage{mathrsfs}

\usepackage{algorithm} 
\usepackage{algpseudocode} 

\usepackage{tensor}
\usepackage{upgreek}
\usepackage{physics}

\usepackage[toc, xindy, acronym]{glossaries}
\usepackage{natbib}

\usepackage[title]{appendix}

\usepackage[setpagesize=false]{hyperref} 
\hypersetup{
	colorlinks,
	linkcolor={blue!80!black},
	citecolor={blue!80!black},
	urlcolor={blue!80!black}
}

\usepackage{enumitem}
\usepackage{zref-savepos}
\usepackage{cleveref}
\crefformat{section}{\S#2#1#3} 
\crefformat{subsection}{\S#2#1#3}
\crefformat{subsubsection}{\S#2#1#3}
\crefformat{figure}{figure~#2#1#3}
\crefformat{appendix}{appendix~#2#1#3}
\crefformat{equation}{(#2#1#3)}

\usepackage{tikz}
\usepackage{circuitikz}
\usetikzlibrary{calc}
\usetikzlibrary{patterns}
\usetikzlibrary{arrows}
\usetikzlibrary{shapes}
\usetikzlibrary{plotmarks}
\usetikzlibrary{decorations.markings}

\usepackage{pgfplots}
\usepgfplotslibrary{polar}
\usepgflibrary{shapes.geometric}
\usepackage[]{pstricks}

\tikzset{
	set arrow inside/.code={\pgfqkeys{/tikz/arrow inside}{#1}},
	set arrow inside={end/.initial=>, opt/.initial=},
	/pgf/decoration/Mark/.style={
		mark/.expanded=at position #1 with
		{
			\noexpand\arrow[\pgfkeysvalueof{/tikz/arrow inside/opt}]{\pgfkeysvalueof{/tikz/arrow inside/end}}
		}
	},
	arrow inside/.style 2 args={
		set arrow inside={#1},
		postaction={
			decorate,decoration={
				markings,Mark/.list={#2}
			}
		}
	},
}

\newcommand{\rmd}{\text{d}}

\ifdefined\partd
\renewcommand{\partd}[3][]{\frac{{\partial^{#1} #2}}{{\partial #3}^{#1}}}
\else
\newcommand{\partd}[3][]{\frac{{\partial^{#1} #2}}{{\partial #3}^{#1}}}
\fi

\renewcommand{\vec}[1]{\boldsymbol{#1}}
\ifdefined\vec  
\renewcommand{\vec}[1]{\boldsymbol{#1}}
\else
\newcommand{\vec}[1]{\boldsymbol{#1}}
\fi

\ifdefined\div
\renewcommand{\div}{{\vec{\nabla} \cdot}}
\else
\newcommand{\div}{{\vec{\nabla} \cdot}}
\fi

\newcommand{\gpar}{\vec{\nabla}_\parallel}
\newcommand{\gperp}{\vec{\nabla}_\perp}

\ifdefined\vr
\renewcommand{\vr}{{\vec{r}}}
\else
\newcommand{\vr}{{\vec{r}}}
\fi

\newcommand{\kperp}{k_\perp}
\newcommand{\kpar}{k_\parallel}

\newcommand{\s}{s}

\newcommand{\vthe}{v_{\text{th}e}}

\newcommand{\rhoi}{\rho_i}
\newcommand{\rhoe}{\rho_e}
\newcommand{\rhos}{\rho_\s}

\newcommand{\dBperp}{{\delta \! \vec{B}_{\!\perp}}}
\newcommand{\Apar}{{A_\parallel}}

\newcommand{\taubar}{\bar{\tau}}

\newcommand{\phik}{{\phi_{\vec{k}}}}
\newcommand{\Apark}{{A_{\parallel \vec{k}}}}

\newcommand{\dne}{\delta n_e}
\newcommand{\uparae}{u_{\parallel e}}
\newcommand{\vphase}{v_\text{ph}}

\newcommand{\phase}{\alpha_{\kperp}}
\newcommand{\energy}{W}
\newcommand{\helicity}{H}
\newcommand{\fluxe}{\varepsilon_\energy}
\newcommand{\fluxh}{\varepsilon_\helicity}
\newcommand{\fluxpm}{\varepsilon^\pm}
\newcommand{\fluxmp}{\varepsilon^\mp}
\newcommand{\fluxp}{{\varepsilon^{+}}}
\newcommand{\fluxm}{{\varepsilon^{-}}}
\newcommand{\forcing}{\sigma_{\varepsilon}}
\newcommand{\imbalance}{\tilde{\sigma}_c}
\newcommand{\uperp}{\vec{u}_\perp}
\newcommand{\thetapm}{{\Theta^\pm}}

\newcommand{\thetap}{{\Theta^+}}
\newcommand{\thetam}{{\Theta^-}}
\newcommand{\thetapmk}{{\Theta_{\vec{k}}^\pm}}
\newcommand{\thetapms}{{\Theta_{\kperp}^\pm}}
\newcommand{\thetamps}{{\Theta_{\kperp}^\mp}}
\newcommand{\thetapk}{{\Theta_{\vec{k}}^+}}
\newcommand{\thetaps}{{\Theta_{\kperp}^+}}
\newcommand{\thetamk}{{\Theta_{\vec{k}}^-}}
\newcommand{\thetams}{{\Theta_{\kperp}^-}}
\newcommand{\tnl}{t_\text{nl}}
\newcommand{\z}{z}
\newcommand{\zed}{z}
\newcommand{\betacrit}{\beta_e^\text{crit}}
\newcommand{\dperp}{D_\perp}
\newcommand{\dpara}{D_\parallel}

\newcommand{\dissratio}{R_\text{diss}}

\title[]{The effects of finite electron inertia on helicity-barrier-mediated turbulence}

\author{T.~Adkins$^{1}$\thanks{Email: toby.adkins@otago.ac.nz}
, 
R.~Meyrand$^{1}$
,
and
J.~Squire$^{1}$
}
\affiliation{
$^1$Department of Physics, University of Otago, Dunedin, 9016, New Zealand
}

\begin{document}  
\maketitle        
	
\begin{abstract}
Understanding the partitioning of turbulent energy between ions and electrons in weakly collisional plasmas is crucial for the accurate interpretation of observations and modelling of various astrophysical phenomena. Many such plasmas are ``imbalanced'', wherein the large-scale energy input is dominated by Alfv\'enic fluctuations propagating in a single direction. In this paper, we demonstrate that when strongly-magnetised plasma turbulence is imbalanced, nonlinear conservation laws imply the existence of a critical value of the electron plasma beta (the ratio of the thermal to magnetic pressures) that separates two dramatically different types of turbulence in parameter space. For betas below the critical value, the free energy injected on the largest scales is able to undergo a familiar Kolmogorov-type cascade to small scales where it is dissipated, heating electrons. For betas above the critical value, the system forms a ``helicity barrier'' that prevents the cascade from proceeding past the ion Larmor radius, causing the majority of the injected free energy to be deposited into ion heating. Physically, the helicity barrier results from the inability of the system to adjust to the disparity between the perpendicular-wavenumber scalings of the free energy and generalised helicity below the ion Larmor radius; restoring finite electron inertia can annul, or even reverse, this disparity, giving rise to the aforementioned critical beta. We relate this physics to the ``dynamic phase alignment'' mechanism (that operates under yet lower beta conditions and in pair plasmas), and characterise various other important features of the helicity barrier, including the nature of the nonlinear wavenumber-space fluxes, dissipation rates, and energy spectra. The existence of such a critical beta has important implications for heating, as it suggests that the dominant recipient of the turbulent energy --- ions or electrons --- can depend sensitively on the characteristics of the plasma at large scales.

\end{abstract}

	
\section{Introduction}
\label{sec:introduction}
Many astrophysical plasma systems are weakly collisional, with their characteristic dynamical timescales approaching those associated with inter-particle collisions. A key question in the context of such plasmas is what determines the partitioning of turbulent free energy between ions and electrons, given that they lack an obvious means of thermal equilibration. Indeed, two-temperature states are expected or observed in a variety of contexts, e.g., accretion discs around black holes \citep{ichimaru77,quataert99}, the intracluster medium \citep{takizawa99,kunz22}, and the solar wind \citep{cranmer09rev}. In the latter context, the Alfv\'enic turbulence launched by low-frequency motions in the corona \citep{depontieu07, tomczyk07} is observed to preferentially heat protons over electrons \citep{hansteen95, cranmer09,bandyopadhyay23}, while heavier minor ions (e.g., helium or oxygen) are heated even more efficiently \citep{kohl97}. This is somewhat puzzling, however, as theories of Alfv\'enic turbulence at low plasma beta (the ratio of the thermal to magnetic pressures) predict that all of the heating from a \cite{K41} style cascade of free energy occurs on electrons \citep{howes08jgr,sch09,howes10,kawazura19,sch19}. Furthermore, such a cascade is unable to easily transfer energy to the small parallel scales required to excite ion-cyclotron waves (ICWs) that can cause efficient perpendicular ion heating \citep{kennel66,isenberg11}. While there are other possible mechanisms that will preferentially heat ions, they either impose constraints on the turbulence that are contradicted by observations or the understanding thereof remains limited. For example, compressive fluctuations are able to cause parallel ion heating \citep{sch09}, but are unlikely to have sufficient power to explain the temperature difference between ions and electrons \citep{howes12}. In a similar vein, mechanisms such as random-walk scattering from ion-Larmor-radius scale electric-field fluctuations \citep[so-called ``stochastic heating'';][]{chandran10} and sub-ion-Larmor-radius kinetic-Alfv\'en-wave (KAW) turbulence \citep{arzamasskiy19,isenberg19} are capable of dissipating a significant fraction of the turbulent energy so long as fluctuations remain of sufficiently large amplitude. Whether solar-wind fluctuations are capable of doing this robustly enough to explain the observed ion heating remains unclear \citep{howes08prl,chandran11}.

A possible explanation for the preferential heating of ions that also explains a number of other solar-wind observations is the so-called ``helicity barrier'' \citep{meyrand21}: when the turbulence is imbalanced (i.e, when there is a significant disparity in the energies of the forwards and backwards propagating fluctuations), as it is in the solar wind, free energy is prevented from cascading past the ion Larmor radius $\rhoi$ and reaching smaller perpendicular scales, where it would, presumably, dissipate on electrons. Instead, the turbulence grows to large amplitudes, creating fine parallel structure that excites ICW fluctuations and heats the ions, which then absorb the majority of the injected power \citep{squire22,squire23}. This helicity-barrier-mediated turbulence has many features that agree with measurements of the low-beta solar wind, including those of the ion velocity distribution function \citep[see, e.g.,][]{marsch06,he15,bowen22}, helicity \citep{huang21,zhao21}, and properties of the steep spectral slopes of the electromagnetic fields in the ``transition range'' on scales comparable to the ion Larmor radius. These transition range spectra have been observed for decades \citep{leamon98,alexandrova09,sahraoui09} and, more recently, by Parker Solar Probe (PSP) \citep{bowen20,duan21,bowen24}. The helicity barrier may also have an important impact on plasmas in other astrophysical contexts, such as in the interpretation of images from the Event Horizon Telescope \citep{wong24}.

This paradigm remains incomplete, however, as research exploring the impact of the helicity barrier in imbalanced solar-wind turbulence has thus far been conducted without accounting for the effects of electron kinetics. In particular, the assumption of a vanishing electron inertial scale $d_e = \rhoe/\sqrt{\beta_e}$ (where $\rhoe$ is the electron Larmor radius and $\beta_e$ is the electron plasma beta) has resulted in the neglect of electron Landau damping, which is particularly significant on scales comparable to $d_e$ \citep{sch09,zocco11,zhou23PNAS}. Given that these effects will undoubtedly play a role in determining the partitioning of turbulent energy between ions and electrons, accounting for them is a necessary extension of the helicity-barrier paradigm to the low-beta regime most relevant to the lower corona. In this paper, we consider the effects of finite electron inertia in imbalanced helicity-barrier-mediated turbulence. Using equations derived in a low-beta asymptotic limit of gyrokinetics, we demonstrate the existence of a critical value of the electron beta $\beta_e$ below which, for a given value of the energy imbalance in the outer-scale fluctuations, the helicity barrier will not form, allowing free energy to cascade to small perpendicular scales. This effect is shown to arise from the constraints placed on the turbulent dynamics by the simultaneous conservation of both free energy and (generalised) helicity across the ion Larmor scale $\rhoi$ and the electron inertial scale $d_e$. At perpendicular scales significantly larger than $\rhoi$ or smaller than $d_e$, both the free energy and the helicity display the same perpendicular-wavenumber scaling, meaning that they can, in principle, cascade simultaneously. This is not the case on scales between $\rhoi$ and $d_e$: below the aforementioned critical beta, the helicity exhibits a shallower scaling with perpendicular wavenumber than the free energy, while above it, it exhibits a steeper scaling. In the former case, fluctuations are able to compensate for this disparity in scaling by becoming increasingly misaligned at small scales through ``dynamic phase alignment'' \citep{loureiro18,milanese20}, allowing the cascade of energy to proceed to small scales uninterrupted. In the latter case, however, fluctuations cannot account for this disparity because they are unable to become more than maximally aligned, eventually leading to the breakdown of the constant-flux solution and the formation of a helicity barrier that prevents energy from cascading past $\rhoi$. The existence of this critical beta thus has important consequences for turbulent heating, acting as a ``switch''  that determines whether the majority of turbulent fluctuations are dissipated on ions (above the critical beta) or electrons (below it) as a function of the equilibrium parameters of the system.

The remainder of this paper is organised as follows. Section \ref{sec:isothermal_KREHM} motivates and outlines the model equations used for theoretical arguments and simulations throughout this work (\cref{sec:fluid_equations}), before considering their linear phase velocity (\cref{sec:linear_waves}) and nonlinearly conserved invariants (\cref{sec:nonlinear_invariants}). Details of the numerical implementation are briefly discussed in \cref{sec:numerics}. Section \ref{sec:imbalanced_turbulence} considers the dynamics of imbalanced turbulence within these model equations. We begin by outlining a theory of balanced turbulence assuming a constant-flux cascade of energy (\cref{sec:constant_flux_cascade}), which exhibits good agreement with numerical simulations. The effect of helicity conservation is then considered in \cref{sec:effect_of_helicity_conservation}, from which the critical value of $\beta_e$ discussed above is shown to follow. The principal characteristics of the helicity barrier are outlined in \cref{sec:the_helicity_barrier} in order to motivate our testing of this critical beta that occurs in \cref{sec:breaking_the_helicity_barrier}, with the prediction being robustly supported by numerical simulations. We then briefly discuss the relationship between the helicity barrier and dynamic phase alignment, before summarising and discussing the implications of our results in \cref{sec:summary_and_discussion}.

\section{Isothermal KREHM}
\label{sec:isothermal_KREHM} 
Standard magnetised plasma turbulence phenomenologies \citep[see, e.g.,][]{GS95, boldyrev06, sch22} imply that fluctuations at scales much smaller than those associated with the variation of the local magnetic field $\vec{B}_0 = B_0 \vec{b}_0$ are highly anisotropic in {space, viz., they satisfy $\kpar \ll \kperp$ for characteristic wavenumbers $\kpar$ and $\kperp$ parallel and perpendicular to $\vec{b}_0$. This anisotropy, which appears to be satisfied in the solar wind \citep{chen13,chen16}, allows even small-scale fluctuations to have frequencies well below the frequency $\Omega_\s$ of the Larmor motion of the particles. Averaging, then, the Vlasov-Maxwell system over this fast Larmor timescale in the presence of such anisotropy leads to the gyrokinetic system of equations \citep{howes06,sch09,abel13}, which has seen recent application to the study of kinetic plasma turbulence in astrophysical plasmas \citep[see, e.g.,][]{howes08prl,howes11prl,kawazura19}. A further simplification can be made by expanding in the limit of low plasma beta, wherein there is minimal coupling between Alfv\'enic and ion-compressive fluctuations because the ion-thermal speed is much smaller than the Alfv\'en one \citep{sch19}. This allows ion kinetics to be neglected even at the ion-Larmor scale. The resultant system of equations is known as the ``kinetic reduced electron heating model'' \citep[KREHM,][]{zocco11,loureiro16viriato}, which couples the equations of reduced magnetohydrodynamics \citep[RMHD,][]{kadomtsev74,strauss76} and electron reduced magnetohydrodynamics \citep[ERMHD,][]{sch09,boldyrev13} to the electron kinetic physics. Although the KREHM equations are formally derived with the electron beta ordered comparable to the electron-ion mass ratio $\beta_e = 8\pi n_{0e} T_{0e}/B_0^2 \sim m_e/m_i$ \citep{zocco11,adkins22}, they remain valid for all $\beta_e \ll 1$ assuming an order-unity equilibrium-temperature-ratio between ions and electrons $\tau \equiv T_{0i}/T_{0e} \sim 1$ ($n_{0e}$ is the equilibrium density of electrons).

In pursuit of further simplicity, we will assume the electrons to be isothermal along exact (equilibrium plus perturbed) field lines. Formally, this amounts to neglecting the parallel gradient of the parallel electron-temperature perturbation that would otherwise appear in the electron momentum equation [see \cref{eq:momentum}]. The effect of this is to decouple the lowest-order fluid moments from the remainder of the kinetic hierarchy (higher-order velocity moments of the kinetic distribution function). Such an approximation is easily justified when investigating dynamics on timescales much slower than the electron parallel-streaming rate, which is typically the case at scales comparable to, or larger than, the ion-Larmor radius $\kperp \rhoi \lesssim 1$ \citep[see, e.g.,][]{sch09,sch19,abel13b,zielinski17}. We note, however, that the typical dynamical timescale associated with the resultant \textit{isothermal KREHM} equations becomes comparable to the electron parallel-streaming rate on scales around the electron-inertial scale $\kperp d_e \sim 1$ (see \cref{sec:linear_waves}). This means that the isothermal approximation breaks down at smaller perpendicular scales, where the effects of electron Landau damping are significant \citep{zhou23PNAS}. Nevertheless --- even though the isothermal approximation cannot be formally derived --- we view the isothermal KREHM equations (which are three-dimensional, having neglected the kinetic physics) as a useful, perhaps even necessary, intermediate step between the simpler RMHD or ERMHD systems and the more complicated KREHM system (which is four-dimensional). In particular, it allows us to investigate the dynamics of the ``fluid'' moments of the system without Landau damping obscuring important features of the dynamics. We will discuss the limitations of the isothermal approximation in detail in \cref{sec:limitations_of_isothermal}.

\subsection{Model equations}
\label{sec:fluid_equations}
The equations of isothermal KREHM are:
\begin{align}
	&\frac{\rmd \dne}{\rmd t}  + n_{0e} \gpar \uparae  =0, \label{eq:continuity}\\
	&m_e n_{0e} \frac{\rmd \uparae}{\rmd t} + T_{0e} \gpar \dne   = - e n_{0e}\left(\frac{1}{c} \frac{\partial A_\parallel}{\partial t} +   \gpar \phi  \right). \label{eq:momentum}
\end{align}
where $\vthe = \sqrt{2T_{0e}/m_e}$ is the thermal speed of electrons, $m_e$ their mass, and $-e$ their charge. Equation \cref{eq:continuity} is the electron continuity equation, describing the advection of the perturbed electron density $\dne$ by the $\vec{E} \times \vec{B}$ flow due to the perturbed electrostatic potential $\phi$:
\begin{align}
	\frac{\rmd}{\rmd  t} = \frac{\partial}{\partial t} + \uperp \cdot \gperp, \quad \uperp = \frac{c}{B_0} \vec{b}_0 \times \gperp \phi, 
	\label{eq:rmdt}
\end{align}
and their compression or rarefaction due to the perturbed parallel electron flow $\uparae$ along the exact magnetic field. The latter includes the perturbation of the magnetic-field direction arising from the parallel component of the magnetic-vector potential $\Apar$:
\begin{align}
	\gpar = \vec{b}\cdot \grad  = \frac{\partial}{\partial z} + \frac{\dBperp}{B_0} \cdot \gperp, \quad \dBperp = - \vec{b}_0 \times \gperp \Apar.
	\label{eq:gpar}
\end{align}
Equation \cref{eq:momentum} is the electron parallel momentum equation, consisting of a balance between electron inertia, the parallel pressure gradient (which, here, is simply the parallel density gradient, due to the assumption of isothermality), and the parallel electric field appearing on the right-hand side. Since an electron flow uncompensated by an ion flow is a current (the ion thermal speed is formally small), $\uparae$ is related to $\Apar$ by Amp\'ere's law:
\begin{align}
	-e n_{0e} \uparae = j_\parallel = \frac{c}{4\pi} \vec{b}_0 \cdot \left( \gperp \times \dBperp\right) \quad \Rightarrow \quad  \uparae = \frac{c}{4\pi e n_{0e}} \gperp^2 \Apar.
	\label{eq:amperes_law}
\end{align}
Finally, the electron-density perturbation is related to $\phi$ by quasineutrality:
\begin{align}
	\frac{\dne}{n_{0e}} = \frac{\delta n_i}{n_{0i}} =  - \taubar^{-1} \frac{e \phi}{T_{0e}} \equiv - \frac{Z}{\tau}(1 - \hat{\Gamma}_0)  \frac{e \phi}{T_{0e}},
	\label{eq:quasineutrality}
\end{align}
where the operator $\hat{\Gamma}_0$ can be expressed, in Fourier space, in terms of the modified Bessel function of the first kind: $\Gamma_0 = I_0(\alpha_i) e^{-\alpha_i}$, where $\alpha_i = (\kperp \rhoi)^2/2$. This becomes $1 - \hat{\Gamma}_0 \approx -\rhoi^2 \gperp^2/2$ at large scales $\kperp \rhoi \ll 1$, and $1 - \hat{\Gamma}_0 \approx 1$ at small scales $\kperp \rhoi \gg 1$; the former limit is why \cref{eq:quasineutrality} is sometimes referred to as the \text{gyrokinetic Poisson equation}. 

Using \cref{eq:amperes_law} and \cref{eq:quasineutrality}, we can write \cref{eq:continuity} and \cref{eq:momentum} as 
\begin{align}
	&\frac{\rmd}{\rmd t} \taubar^{-1} \frac{e\phi}{T_{0e}} - \frac{c}{4 \pi e n_{0e}} \gpar \gperp^2 \Apar = 0,  \label{eq:phi_equation} \\
	&\frac{\rmd}{\rmd t}\left(\Apar - d_e^2 \gperp^2 \Apar\right) = - c \left(\frac{\partial \phi}{\partial z} + \gpar \taubar^{-1} \phi\right). \label{eq:apar_equation}
\end{align}
Together, \cref{eq:phi_equation} and \cref{eq:apar_equation} form a closed pair of equations describing the evolution of the electromagnetic potentials $\phi$ and $\Apar$ of a strongly-magnetised, low-beta plasma in the absence of electron Landau damping. At scales $\kperp d_e \ll 1$, we recover from \cref{eq:phi_equation} and \cref{eq:apar_equation} the FLR-MHD system studied in \cite{meyrand21}, which itself reduces to RMHD and ERMHD at large ($\kperp \rhoi \ll 1$) and small ($\kperp \rhoi \gg 1$) scales, respectively. In the ultra-low beta limit $\beta_e \ll m_e/m_i$, in which the electron-inertial length becomes larger than the ion-Larmor radius $d_e \gg \rhoi$, we also recover equations describing inertial-Aflv\`en-wave turbulence considered by \cite{loureiro18} and \cite{milanese20}. To re-iterate, \cref{eq:phi_equation} and \cref{eq:apar_equation} can be simply obtained from the KREHM system (which itself can be derived from gyrokinetics in the low-beta limit) by neglecting higher-order moments of the electron kinetic distribution function. Thus, our model equations share all of KREHM's physical characteristics apart from electron Landau damping.

\subsection{Phase velocity}
\label{sec:linear_waves} 
Linearising and Fourier-transforming \cref{eq:phi_equation} and \cref{eq:apar_equation}, we find forwards and backwards propagating modes of frequency $\omega = \pm \kpar \vphase (\kperp)$, where the perpendicular-wavenumber-dependant phase velocity is given by
\begin{align}
	\vphase(\kperp) =  \kperp \rhos \left(\frac{1 + \taubar}{1 + \kperp^2 d_e^2}\right)^{1/2}v_A.
	\label{eq:vphase}
\end{align}
Here, $\rhos = \sqrt{Z/2\tau} \rhoi$ is the ion-sound radius, related to the (thermal) sound speed by $c_s = \rhos/\Omega_i$, and $v_A = B_0/\sqrt{4\pi n_{0i} m_i}$ is the Alfv\'en speed. Equation \cref{eq:vphase} has the asymptotic behaviour 
\begin{align}
	\lim_{\kperp \rightarrow \: 0} \vphase(\kperp) = v_A, \quad 	\lim_{\kperp \rightarrow \: \infty} \vphase(\kperp) =\left(\frac{1+\tau/Z}{2}\right)^{1/2} \vthe,
	\label{eq:vphase_limits} 
\end{align}
meaning that we recover Alfv\'en waves and the electron parallel-streaming rate at the largest and smallest scales, respectively. Its behaviour in the intermediate region, however, depends on the relative sizes of the ion-sound radius $\rhos$ and electron-inertial scale $d_e$, a competition that is controlled (ignoring any $\tau$ or $Z$ dependence) by the ratio of the electron beta to the electron-ion mass ratio $\beta_e (m_e/m_i)^{-1}$. In particular, \cref{eq:vphase} is an \textit{increasing} function of (perpendicular) wavenumber for $\beta_e \gg m_e/m_i$, and a \textit{decreasing} one for $\beta_e \ll m_e/m_i$, viz., 
\begin{align}
	\vphase(\kperp) \propto 
	\left\{
		\begin{array}{cc}
				\displaystyle \kperp \rhos , & \beta_e \gg m_e/m_i, \\ [4mm]
				\displaystyle (\kperp d_e)^{-1}, &  \beta_e \ll m_e/m_i, 
			\end{array}	 
		\right.
		\label{eq:vphase_scalings}
\end{align}
for wavenumbers in the intermediate region. The associated waves are known as kinetic Alfv\'en waves (KAWs), for $\beta_e \gg m_e/m_i$, and inertial Alfv\'en waves, for $\beta_e \ll m_e/m_i$. This behaviour of the phase velocity is manifest in \cref{fig:vphase}, where we plot \cref{eq:vphase} for different values of $\beta_e$.\footnote{It is worth clarifying that $\beta_e$ itself is not a parameter of the isothermal KREHM system of equations \cref{eq:phi_equation} and \cref{eq:apar_equation} due to the fact that they are asymptotically derived under the low-beta ordering $\beta_e \sim m_e/m_i$. This means that, formally, $\beta_e$ should only appear when normalised to the electron-ion mass ratio, viz., $\beta_e(m_e/m_i)^{-1}$ is the true parameter that we will vary throughout this paper. On a similar note, the isothermal KREHM system is derived assuming that $\kperp \rhoi \sim 1$, meaning that we are (formally) allowed to make $\kperp \rhoi$ as large as we would like.~Physically, however, the isothermal KREHM system only applies to wavenumbers $\kperp \lesssim \rhoe^{-1}$, and so dynamics at scales $\kperp \rhoi \gtrsim \sqrt{m_e/m_i}$ are only valid in an asymptotic sense.} Note that for $\beta_e \sim m_e/m_i$, the curve is non-monotonic; see inset of \cref{fig:vphase}. Crucially, the increase, or otherwise, of \cref{eq:vphase} with $\kperp$ determines when the helicity barrier must form, and so whether or not a significant fraction of energy is able to cascade towards small scales; this is discussed in \cref{sec:effect_of_helicity_conservation}.

\begin{figure}
	
	\includegraphics[width=1\textwidth]{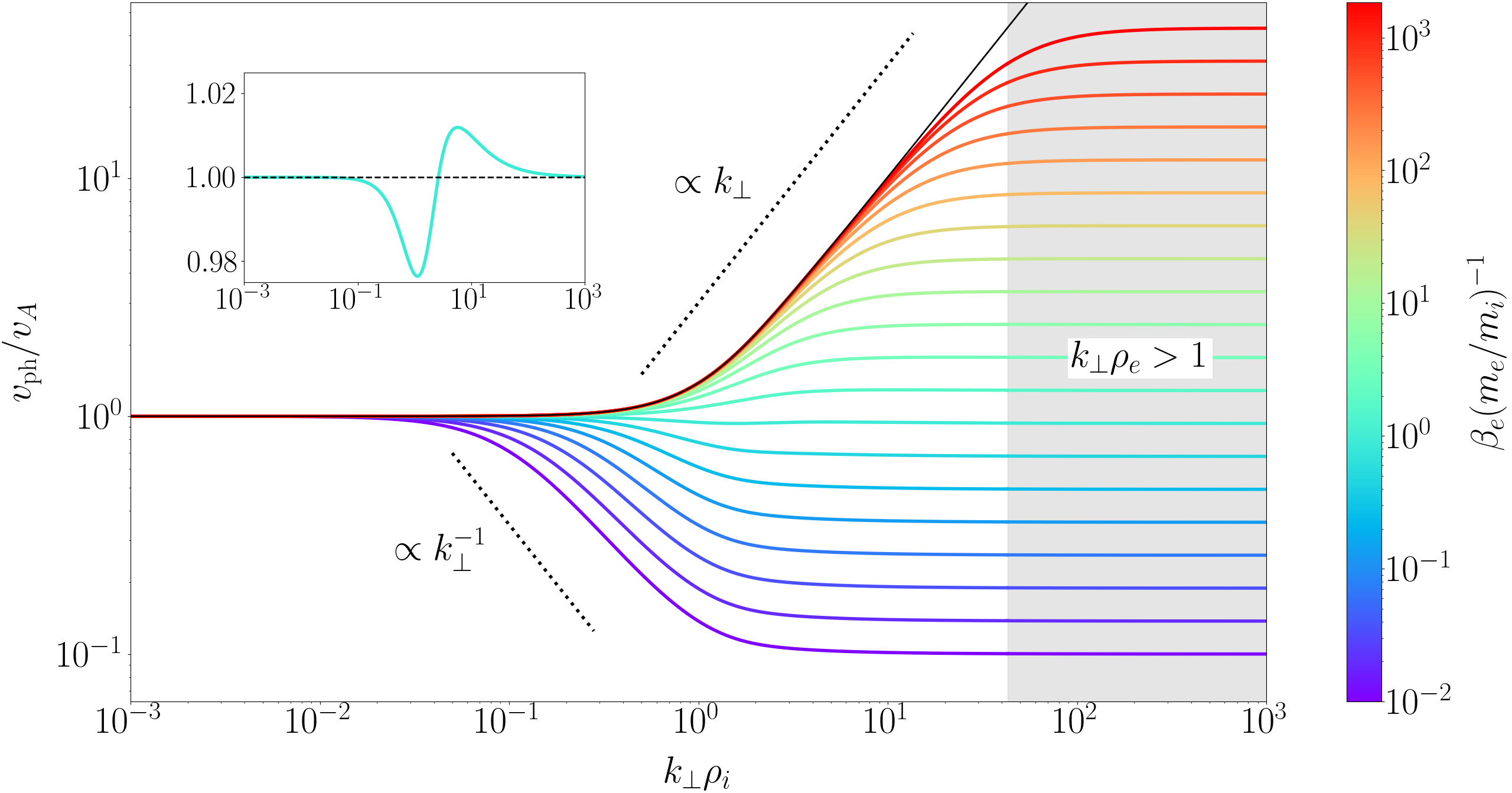}
	
	\caption[]{The phase velocity \cref{eq:vphase}, normalised to the Alfv\'en speed, plotted as a function of perpendicular wavenumber $\kperp \rhoi$, and for $\tau = Z =1$. The colours indicate the value of $\beta_e (m_e/m_i)^{-1}$ for a given line, with the solid black line showing the FLR-MHD case ($d_e \rightarrow 0$). The dotted lines are the scalings \cref{eq:vphase_scalings}. The vertical shaded region indicates wavenumbers $\kperp \rho_e > 1$ for which the model ceases to apply. The inset panel shows the case of $\beta_e = m_e/m_i$, with the horizontal dashed line indicating $\vphase/v_A = 1$.}
	\label{fig:vphase}
\end{figure}

The eigenfunctions associated with the forwards and backwards propagating modes can be expressed, in Fourier space, as:
\begin{align}
	\thetapmk \equiv \sqrt{1 + \kperp^2 d_e^2 }\left[\frac{c}{B_0} \frac{\vphase(\kperp)/v_A}{(\kperp \rhos)^2} \taubar^{-1} \phik  \mp \frac{\Apark}{\sqrt{4 \pi n_{0i}  m_i}}\right].
	\label{eq:thetapm}
\end{align}
Apart from the prefactor multiplying the square brackets and the difference in $\vphase(\kperp)$, this definition is identical to that adopted in \cite{meyrand21}. These generalised Els\"asser potentials have the property that on the largest scales $\kperp \ll \rhoi^{-1}, d_e^{-1}$, they reduce to the standard RMHD Els\"asser potentials  \citep{elsasser50}, viz.,
\begin{align}
	\lim_{\kperp \rightarrow \: 0} \vec{b}_0 \times \gperp \Theta^\pm = \vec{\zed}^\pm \equiv \uperp \pm  \frac{\dBperp}{\sqrt{4\pi n_{0i} m_i}}.
	\label{eq:thetapm_rmhd}
\end{align}
These potentials \cref{eq:thetapm} provide a natural basis for our investigation of imbalanced turbulence in isothermal KREHM.

\subsection{Nonlinear invariants}
\label{sec:nonlinear_invariants}
Most turbulent systems possess at least one nonlinear invariant --- a quantity that is conserved by nonlinear interactions but may have localised sources (e.g., forcing or equilibrium gradients) and sinks (e.g., viscosity or particle collisions). Gyrokinetics conserves the so-called \textit{free energy}, which is the sum of quadratic norms of the magnetic perturbations, and the perturbations of the distribution functions of both ions and electrons. In the isothermal KREHM system \cref{eq:phi_equation} and \cref{eq:apar_equation}, the free energy takes the form \citep{zocco11,loureiro16viriato,adkins22}: 
\begin{align}
	\energy & = \int \frac{\rmd^3 \vec{r}}{V} \left[\frac{e^2 n_{0e}}{2 T_{0e}} \left(\phi \taubar^{-1} \phi\right) + \frac{e^2 n_{0e}}{2 T_{0e}} \left(\taubar^{-1} \phi \right)^2 + \frac{ \left| \gperp \Apar\right|^2 + d_e^2 \left(\gperp^2 \Apar \right)^2}{8 \pi} \right],
	\label{eq:free_energy} 
\end{align}  
where $V$ is the plasma volume.
The contributions to \cref{eq:free_energy} are, from left to right, the energies associated with perturbations of the electrostatic potential, electron density, (perpendicular) magnetic field, and electron parallel-velocity. At large scales $\kperp \ll \rhoi^{-1}, d_e^{-1}$, this becomes
\begin{align}
	W \approx \frac{n_{0i} m_i}{2} \int \frac{\rmd^3 \vec{r}}{V} \left(\left| \uperp \right|^2 + \frac{\left|\dBperp\right|^2}{4 \pi n_{0i} m_i}\right) = \frac{n_{0i} m_i}{4} \int \frac{\rmd^3 \vec{r}}{V} \left( \left|\vec{\zed}^+\right|^2 + \left|\vec{\zed}^-\right|^2  \right),
	\label{eq:free_energy_rmhd}
\end{align} 
recovering the usual expression for the free energy in RMHD \citep[see, e.g.,][]{sch09}.

Free energy is normally the quantity whose cascade from large (injection) to small (dissipation) scales determines the properties of the plasma's turbulent state \citep{sch08,sch09}, as in hydrodynamic turbulence \citep[see, e.g.,][and references therein]{alexakis18}. Isothermal KREHM, however, possesses another invariant that also plays an important role --- the \textit{generalised helicity}:
\begin{align}
	\helicity = -\frac{e^2 n_{0e} v_{A}}{cT_{0e}} \int \frac{\rmd^3 \vec{r}}{V} \: \taubar^{-1}\phi  \left(\Apar- d_e^2 \gperp^2 \Apar \right).
	\label{eq:helicity}
\end{align} 
This reduces to the MHD cross-helicity at $\kperp \ll \rhoi^{-1}, d_e^{-1}$, viz.,
\begin{align}
	H \approx n_{0i} m_i \int \frac{\rmd^3 \vec{r}}{V} \:  \frac{\uperp \cdot \dBperp}{\sqrt{4 \pi n_{0i} m_i}} = \frac{n_{0i} m_i}{4} \int \frac{\rmd^3 \vec{r}}{V} \left( \left|\vec{\zed}^+\right|^2 - \left|\vec{\zed}^-\right|^2  \right) ,
	\label{eq:helicity_rmhd}
\end{align}
and is proportional to the magnetic helicity at $\rhoi^{-1} \ll \kperp \ll d_e^{-1}$ (due to perpendicular pressure balance; see \citealt{sch09}). The presence of this generalised helicity places an additional constraint on the dynamical states accessible by the system, as the turbulence must now evolve in such a way as to conserve both \cref{eq:free_energy} and \cref{eq:helicity} simultaneously. The remainder of this paper is devoted to demonstrating how and when these constraints give rise to different turbulent states, and the consequences that this may have for plasma heating. 

\subsection{Numerical setup}
\label{sec:numerics}
In what follows, the isothermal KREHM system \cref{eq:phi_equation}-\cref{eq:apar_equation} is solved using a modified version of the pseudospectral code TURBO \citep{teaca09} in a triply-periodic box of size $L_x = L_y = L_\z = L$ with $n_\perp^2 \times n_\z$ Fourier modes. A third-order \cite{williamson80} algorithm is used for the time-stepping. Time is measured in units of the parallel Alfv\'en time $t_A = L_\z/v_A$. Hyper-dissipation in both the perpendicular and parallel directions is introduced by replacing the time-derivative of the left-hand sides of \cref{eq:phi_equation} and \cref{eq:apar_equation} by 
\begin{align}
	\frac{\rmd}{\rmd t} + \nu_\perp \gperp^6 + \nu_\z \frac{\partial^6}{\partial z^6}.
	\label{eq:hyperdissipation}
\end{align}
The coefficients $\nu_\perp$ and $\nu_\z$ are adaptive, viz., they are re-evaluated at each timestep to ensure that dissipation occurs near the grid scale, maximising the inertial range \citep[details of the numerical implementation can be found in][]{meyrand24}. Fluctuations are forced at large scales at $\kperp = 4\pi/L$, $|k_\z| = 2\pi/L$ through the form of negative damping \citep{meyrand21}; this method allows the rates of free-energy and helicity injection to be controlled exactly while producing sufficiently random motions to generate turbulence. All of the simulations shown in \cref{tab:simulation_parameters} have $\tau = Z = 1$, though we will retain dependencies on these parameters in analytical expressions for the sake of completeness.

\begin{table}
	
	\centering
	
	\begin{tabular}{l | l  c c c c c c }
		\multicolumn{2}{c}{} & Resolution & $\forcing$ & $\rhoi/L$ & $d_e/L$ &$\beta_e (m_e/m_i)^{-1}$ & Sims \vspace*{-1mm}\\
		\hline
		High resolution & CF-res & $1024^3$  & 0.80 & 0.100 & 0.100 & 1.00 & 1 \\
		&  HB-res & $1024^3$  & 0.80 & 0.100 & 0.050 & 4.00 & 1  \\
		& RMHD & $1024^3$  & 0.80 & -     & -     & -    & 1 \\
		\hline
		Comparison & CF  & $256^3$  & 0.80 & 0.100 & 0.100 & 1.00 & 1 \\
		& HB  & $256^3$  & 0.80 & 0.100 & 0.050 & 4.00 & 1  \\
		& ULB & $256^3$  & 0.80 &   -   & 0.100 & -    & 1  \\
		\hline
		Beta scan &  & $256^3$ & 0.20 & 0.100 & (0.004, 0.110) & (0.83, 625.00) & 10 \\
		&  & $256^3$ & 0.30 & 0.100 & (0.015, 0.110) & (0.83, 44.40) & 8 \\
		&  & $256^3$ & 0.40 & 0.100 & (0.015, 0.110) & (0.83, 44.40) & 8 \\
		&  & $256^3$ & 0.50 & 0.100 & (0.015, 0.110) & (0.83, 44.40) & 8 \\
		&  & $256^3$ & 0.60 & 0.100 & (0.040, 0.110) & (0.83, 6.25)  & 8 \\
		&  & $256^3$ & 0.70 & 0.100 & (0.040, 0.110) & (0.83, 6.25)  & 8 \\
		&  & $256^3$ & 0.80 & 0.100 & (0.040, 0.110) & (0.83, 6.25)  & 8 \\
		&  & $256^3$ & 0.90 & 0.100 & (0.060, 0.110) & (0.83, 2.78)  & 6 \\
		&  & $256^3$ & 0.99 & 0.100 & (0.100, 0.110) & (0.83, 1.00)  & 2 \\
		\hline
		Resolution scan &  & $64^3$  & (0.55, 0.65) & 0.100 & - & - & 3 \\
		&  & $128^3$ & (0.25, 0.35) & 0.100 & - & - & 3 \\
		&  & $192^3$ & (0.15, 0.25) & 0.100 & - & - & 3 \\
		&  & $256^3$ & (0.10, 0.20) & 0.100 & - & - & 3 \\
		\hline 
	\end{tabular}

	\caption{The parameters used for the isothermal KREHM simulations considered in this paper. All simulations have $\tau = Z = 1$. Values in parentheses indicate the minimum and maximum values for the corresponding column, with the final column (`sims') indicating the number of simulations in a given set. A dash in an entry indicates that the physical simulation being considered does not contain that physical parameter.}
	\label{tab:simulation_parameters}
	
\end{table} 

\section{Imbalanced Alfv\'enic turbulence}
\label{sec:imbalanced_turbulence}
Observations show that solar-wind turbulence is imbalanced, meaning it is energetically dominated by outward propagating Alfv\'enic structures associated with $\vec{\zed}^+$ (inward propagating structures are associated with $\vec{\zed}^-$). It must, therefore, possess a non-zero cross-helicity \cref{eq:helicity_rmhd}. Writing the free energy \cref{eq:free_energy} and (generalised) helicity \cref{eq:helicity} in terms of the generalised Els\"asser potentials \cref{eq:thetapm} as 
\begin{align}
	\energy & = \frac{n_{0i} m_i}{4} \sum_{\vec{k}} \left( \left|\kperp\thetapk\right|^2 + \left|\kperp \thetamk\right|^2 \right), \label{eq:free_energy_thetapm} \\
	 \helicity & = \frac{n_{0i} m_i}{4} \sum_{\vec{k}} \frac{ \left|\kperp\thetapk\right|^2 - \left|\kperp \thetamk\right|^2 }{\vphase(\kperp)/v_A},
	\label{eq:helicity_thetapm}
\end{align}
it is clear that the same will be true of the isothermal KREHM system given a difference in the $\thetapm$ energies. We quantify this \textit{energy imbalance} by the ratio of the free energy to the helicity:
\begin{align}
	\imbalance = \frac{H}{W}, 
	\label{eq:imbalance}
\end{align}
which reduces to the normalised cross-helicity (or RMHD imbalance) $\sigma_c$ at large scales: 
\begin{align}
	\lim_{\kperp \rightarrow \: 0} \imbalance = \sigma_c = \frac{\int \rmd^3 \vec{r} \left( \left|\vec{\zed}^+\right|^2 - \left|\vec{\zed}^-\right|^2  \right)}{\int \rmd^3 \vec{r} \left( \left|\vec{\zed}^+\right|^2 + \left|\vec{\zed}^-\right|^2  \right)} \leqslant 1.
	\label{eq:imbalance_limit}
\end{align}
Measured values of \cref{eq:imbalance_limit} often exceed $|\sigma_c| \gtrsim 0.8$ in the solar-wind, particularly in near-sun regions \citep{mcmanus20}. Despite this, a comprehensive theory of imbalanced turbulence remains elusive, even in the (simpler) context of RMHD \citep{perez09,chandran08,beresnyak09,lithwick07,chandran19,sch22}. As such, the phenomenological theory presented in the following sections lays no claim to being comprehensive; instead, it should be viewed as a useful framework through which the effect of finite electron inertia can be explored in imbalanced turbulence. 

\subsection{Constant-flux cascade}
\label{sec:constant_flux_cascade} 
Consider the case where (free) energy \cref{eq:free_energy} and (generalised) helicity \cref{eq:apar_equation} are injected into our isothermal KREHM system at constant rates $\fluxe$ and $\fluxh$, respectively, by some large-scale stirring of turbulent fluctuations (due to, e.g., reflection of outwards propagating fluctuations; \citealt{velli89}). We denote the resultant \textit{injection imbalance} --- the ratio of the injected flux of helicity to that of the energy --- as $\sigma_\varepsilon = |\fluxh|/\fluxe$. Given that both the energy and helicity are nonlinear invariants, the only available route to their dissipation is through some nonlinear transfer to small scales. Motivated by this, we assume, for the moment, that there is a local, \cite{K41} style cascade that carries a constant flux of injected energy and helicity from the outer (injection) scale, through some putative inertial range, to the dissipation scale. 
 
It follows immediately from this that the rates of energy injection into the forward and backward propagating fluctuations 
\begin{align}
	\fluxpm = \frac{\fluxe \pm \fluxh}{2} =\frac{1 \pm \forcing}{2}\fluxe,
	\label{eq:fluxpm}
\end{align}
will be equal to the associated flux of $\thetapm$ energy through the inertial range. We estimate these energy fluxes from \cref{eq:free_energy_thetapm} as:
\begin{align}
	\frac{1}{n_{0e} T_{0e}} \frac{\rmd W_\pm}{\rmd t} \sim \left(\tnl^\pm \right)^{-1} \frac{\left(\kperp \thetapms \right)^2}{c_s^2} \sim \fluxpm = \text{const},
	\label{eq:fluxpm_definition}
\end{align}
where here, and in what follows, $\thetapms$ refers to the characteristic amplitude of the Els\"asser potentials at the scale $\kperp^{-1}$, rather than to the Fourier transform of the field \cref{eq:thetapm}. Formally, $\thetapms$ can be defined by
\begin{align}
	\frac{\left(\kperp \thetapms \right)^2}{c_s^2} = \frac{1}{n_{0e}T_{0e}}\int_{\kperp}^\infty \rmd \kperp' \: E_\perp^\pm (\kperp') , \quad E_\perp^\pm(\kperp) = 2\pi \kperp  \int_{-\infty}^\infty \rmd \kpar \:  \frac{n_{0i}m_i}{4}\left< \left| \kperp \thetapmk \right|^2 \right>,
	\label{eq:thetapm_spectra_def}
\end{align}
where $E_\perp^\pm(\kperp)$ is the 1D perpendicular energy spectrum of $\thetapm$ [cf. \cref{eq:free_energy_thetapm}], and in which the brackets denote an ensemble average. An alternative definition would be via a second-order structure function \citep[see, e.g.,][]{davidson13}. Perturbations of other quantities, such as potential, velocity, magnetic field, etc., will similarly be taken to refer to their characteristic amplitude at a given perpendicular scale. 

In order to proceed, we need an expression for the nonlinear times appearing in \cref{eq:fluxpm_definition}. However, as we will discuss shortly, determining exactly what these nonlinear times are is not a straightforward task, remaining an open research question even in the RMHD regime \citep{sch22}. As such, let us henceforth consider the balanced regime, assuming that the rates of energy injection into the backwards and forwards propagating fluctuations are comparable, and that they have the same scaling with perpendicular wavenumber, such that their ratio is constant at all scales, viz.,
\begin{align}
	\fluxp \sim \fluxm \sim \fluxe, \quad \frac{\thetaps}{\thetams} = \text{const}.
	\label{eq:balanced_assumptions}
\end{align}
Then, the nonlinear times in \cref{eq:fluxpm_definition} can straightforwardly be taken to be the nonlinear $\vec{E} \times \vec{B}$ advection rate associated with either field, which, comparing \cref{eq:rmdt} and \cref{eq:thetapm}, and neglecting any possible anisotropy in the perpendicular plane, can be written as  
\begin{align}
	\left(\tnl^\pm \right)^{-1} \sim \kperp u_\perp \sim \Omega_i \left(\frac{\taubar^2}{1+\taubar}\right)^{1/2} \left(\kperp \rhos\right)^3 \left(\frac{\thetamps}{\rhos c_s}\right).
	\label{eq:nonlinear_times}
\end{align}
Combining \cref{eq:fluxpm_definition}, \cref{eq:nonlinear_times}, and \cref{eq:balanced_assumptions}, we obtain an estimate for the fluctuations of the Els\"asser potentials:
\begin{align}
	\frac{\thetapms}{\rhos c_s} \sim \left(\frac{\fluxe}{\Omega_i}\right)^{1/3}\left(\frac{1+\taubar}{\taubar^2}\right)^{1/6} \left(\kperp \rhos\right)^{-5/3},
	\label{eq:thetapm_scaling}
\end{align}
and the associated spectra:
\begin{align}
	E_\perp^\pm(\kperp)  \propto \left(\frac{1+
		\taubar}{\taubar^2}\right)^{1/3} \left(\kperp \rhos\right)^{-7/3} \propto \left\{
	\begin{array}{cc}
		\displaystyle \kperp^{-5/3} , & \kperp \rhoi \ll 1, \\ [4mm]
		\displaystyle \kperp^{-7/3}, &  \kperp \rhoi \gg 1.
	\end{array}	 
	\right.
	\label{eq:thetapm_spectra}
\end{align}
These are the standard Kolmogorov-style scalings for both RMHD and KAW turbulence, respectively \citep[see, e.g.,][]{GS95,cho04,sch09}, which hold even in the presence of finite electron inertia [only the magnetic energy exhibits a transition at $\kperp d_e \sim 1$; see \cref{eq:magnetic_spectrum}]. However, simulations of forced RMHD turbulence have consistently shown scaling exponents closer to $-3/2$, while those of KAW turbulence appear to be closer to $-8/3$. The departure of both exponents from those derived here has been shown to arise due to intermittency effects (see, e.g., \citealt{boldyrev06,chandran15,mallet17a}, in the former case, or \citealt{boldyrev12,meyrand13,zhou23MNRAS}, in the latter), which we have not attempted to account for here. Given that our aim is not a comprehensive theory of turbulence, but instead to highlight the effect of helicity conservation in such an environment, we consider agreement with either the derived spectra \cref{eq:thetapm_spectra} or those corrected for intermittency to be sufficient evidence that the system is undergoing a constant-flux cascade. We will compare with a scaling exponent of $-3/2$ where necessary as it is well-motivated in balanced turbulence and supported by observations of imbalanced turbulence \citep{chen20}.

Perhaps the most important feature not captured by the scalings \cref{eq:thetapm_scaling} and \cref{eq:thetapm_spectra} is the imbalance, viz., the difference in amplitudes between each of the fields: we have assumed that $\thetaps \sim \thetams$, which clearly will not be the case if $\fluxp \gg \fluxm$. One way of rectifying this would be to relax the first assumption in \cref{eq:balanced_assumptions} and assume that the relevant nonlinear time for each field is the $\vec{E} \times \vec{B}$ advection rate associated with the \textit{counter-propagating} field, which yields, via a straightforward generalisation of the argument above, 
	\begin{align}
		\thetapms \propto \left[\frac{\left(\fluxpm\right)^2}{\fluxmp}\right]^{1/3} \quad \Rightarrow \quad	\frac{\thetaps}{\thetams} \sim \frac{\fluxp}{\fluxm} = \text{const}.
		\label{eq:field_ratio}
	\end{align}
This is the conclusion at which \cite{lithwick07} arrived in the context of strongly-imbalanced RMHD turbulence. Though there is numerical evidence to suggest that this approximately holds for the ratio of their associated energies \citep{beresnyak08,beresnyak09slopes,beresnyak10,mallet11,sch22}, evidence for it holding throughout the inertial range is less clear. Previous work \citep[see, e.g.,][]{perez10a,mallet11} suggests that the stronger field typically has a steeper spectrum than the weaker one, although this difference in their slopes tends to decrease as numerical resolution is increased. Another potential issue following from \cref{eq:field_ratio} is that the ratio of the nonlinear times for each field must scale as $\tnl^+/\tnl^- \sim \fluxp/\fluxm$ [this follows from combining \cref{eq:nonlinear_times} and \cref{eq:field_ratio}]. As pointed out by \cite{lithwick07}, this has the counterintuitive implication that the weaker $\thetams$ perturbation, which is advected by $\thetaps$ at a faster rate $(\tnl^-)^{-1}$, can nevertheless coherently advect $\thetaps$ at the slower rate $(\tnl^+)^{-1}$. Though they propose a disparity between the spatial and temporal coherences of each field as a possible explanation for this, such a state is hard to justify in general; for a possible alternative explanation, see section 9.6 of \cite{sch22}. Finally, while the assumption that the counter-propagating field is the only source of nonlinear advection is guaranteed to be satisfied in the RMHD regime ($\kperp \rhoi \ll 1$), this is not obviously true at sub-ion scales ($\kperp \rhoi \gg 1$). The dispersive nature of KAWs makes nonlinear interactions between co-propagating perturbations (i.e., $\thetapm$ with $\thetapm$) possible at these scales, meaning that it is, in principle, possible to support a turbulent cascade with a single component $\thetapm$ \citep{cho11,kim15,voitenko16}. Given, however, that these subtleties take us beyond the main focus of this work, we will not engage further with them here, having highlighted our reasons for presenting only a balanced turbulence phenomenology when deriving the scaling predictions \cref{eq:thetapm_spectra}.

\begin{figure}
	
	\centering
	
	\begin{tabular}{c}
		
		\includegraphics[width=1\textwidth]{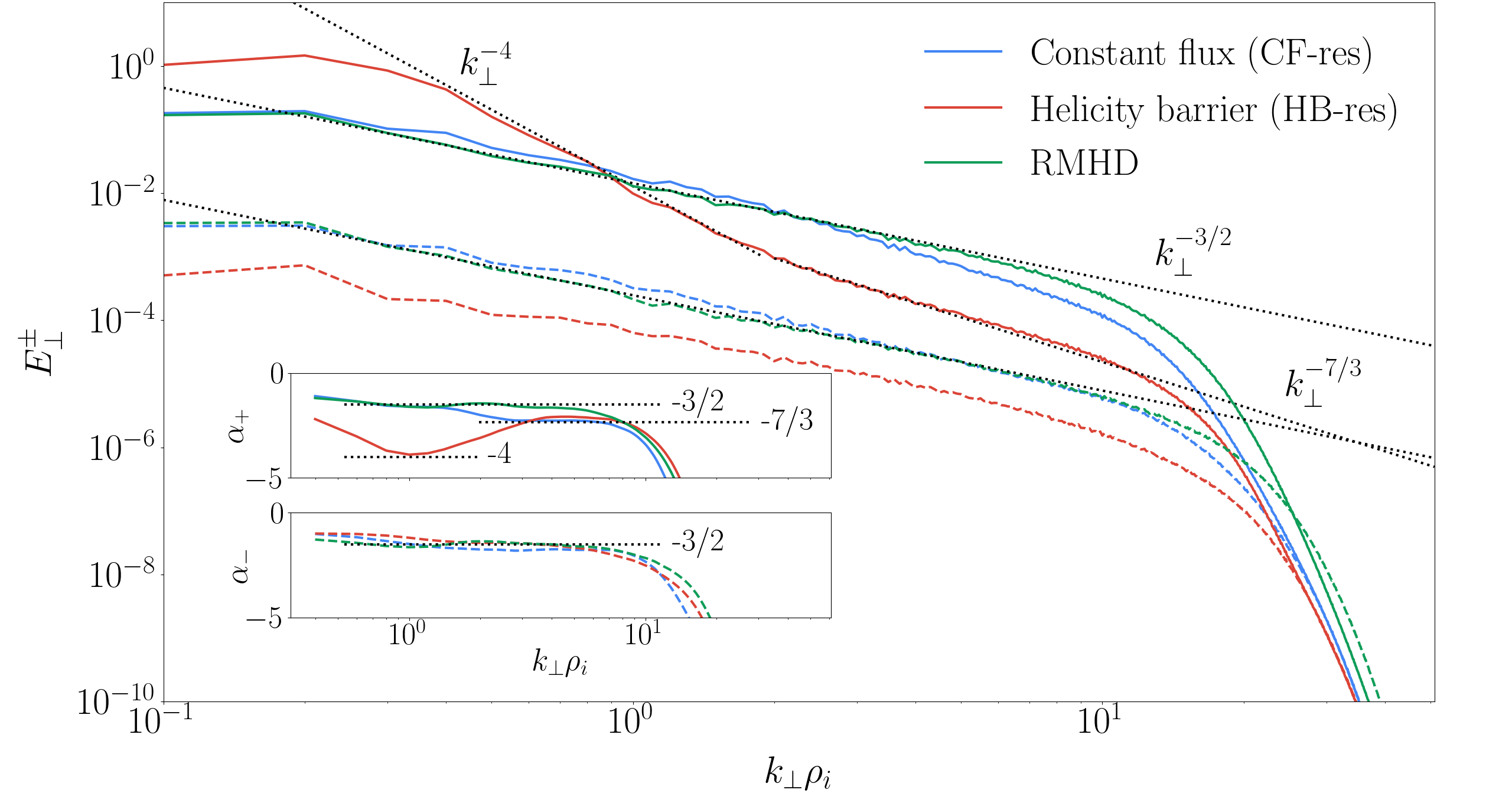}  \\\\
		(a) $\thetapm$ energy spectra \\\\
		\includegraphics[width=1\textwidth]{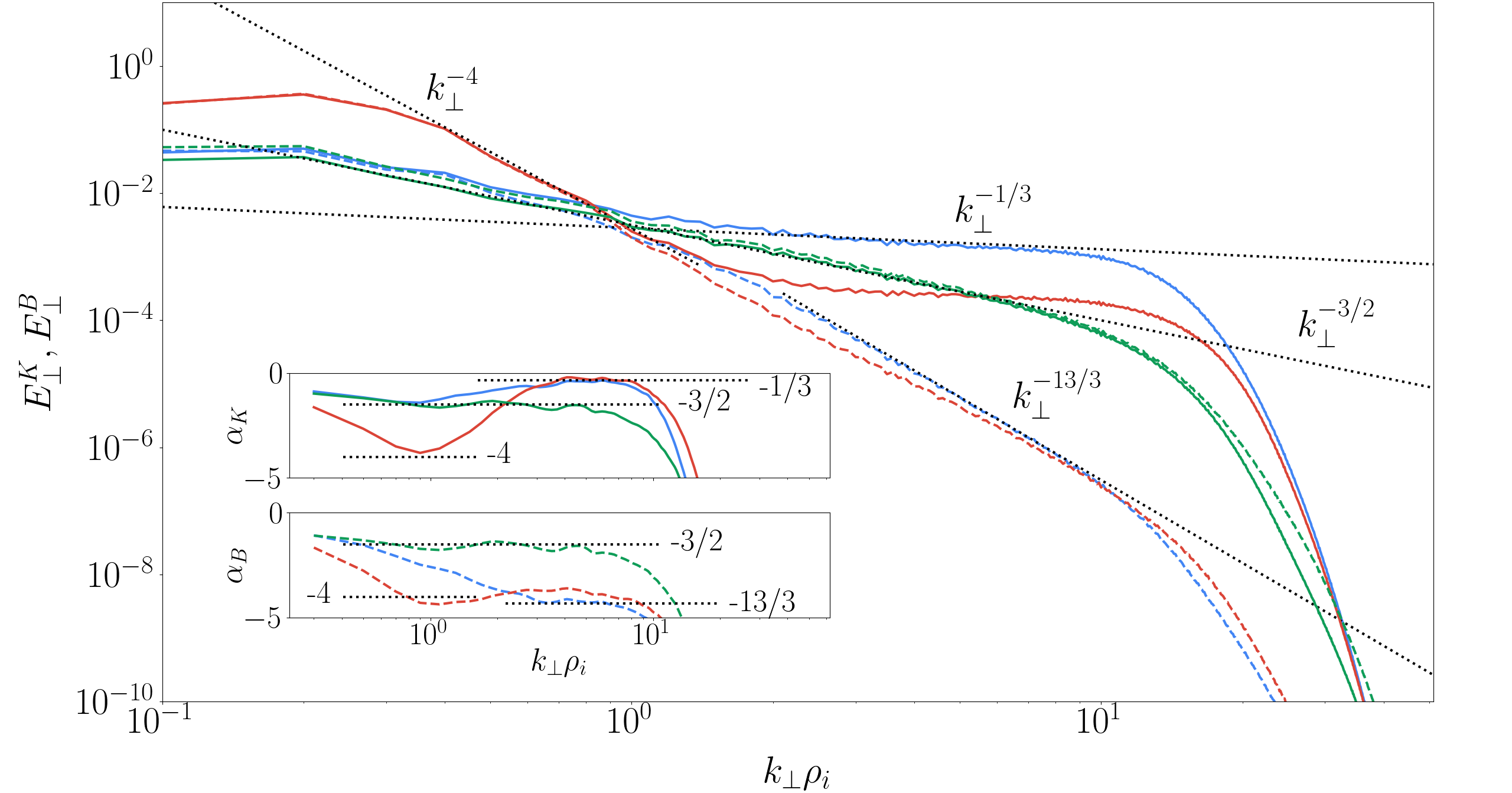}  \\\\
		(b) Kinetic and magnetic energy spectra
		
	\end{tabular}
	
	\caption[]{One-dimensional perpendicular energy spectra for the ``high resolution'' simulations in \cref{tab:simulation_parameters}: CF (blue), HB (red), and RMHD (green). Solid and dashed lines correspond to (a) $E_\perp^+(\kperp)$ and $E_\perp^-(\kperp)$ or (b) $E_\perp^K(\kperp)$ and $E_\perp^B(\kperp)$, respectively. The insets panels show the local scaling exponents $\alpha_{(\dots)} = \rmd \log E^{(\dots)}_\perp/\rmd \log \kperp$ for each spectrum. The inertial-range scalings \cref{eq:thetapm_scaling}, \cref{eq:kinetic_spectrum}, and \cref{eq:magnetic_spectrum} are shown by black dotted lines [with $-5/3$ scalings replaced with $-3/2$, as discussed following \cref{eq:thetapm_spectra}]. Note that the axis of the (scale invariant) RMHD simulation has been rescaled for comparison with the other two cases. Both simulations CF (constant flux) and RMHD are expected to saturate via a constant-flux cascade, and show good agreement with the predicted scalings, while simulation HB (helicity barrier) exhibits entirely different scalings.}
	\label{fig:refined_spectra}
\end{figure}

For completeness, we include the scalings of the two contributions to the free energy at RMHD scales [see \cref{eq:free_energy_rmhd}]: the kinetic energy associated with the $\vec{E} \times \vec{B}$ flow $\propto |\uperp|^2$, and the energy contained in perpendicular magnetic field fluctuations $\propto |\dBperp|^2$. Defining their spectra, analogously to \cref{eq:thetapm_spectra_def}, as
\begin{align}
	E_\perp^K(\kperp) &= 2\pi \kperp\int_{-\infty}^\infty \rmd \kpar \:  \frac{e^2 n_{0e}}{2 T_{0e}} \rhos^2 \left<  |\kperp \phik|^2 \right>, \label{eq:kinetic_spectrum_def} \\
	E_\perp^B(\kperp) &= 2\pi \kperp\int_{-\infty}^\infty \rmd \kpar \: \frac{\left<  |\kperp \Apark|^2 \right>}{8\pi} ,
	\label{eq:magnetic_spectrum_def}
\end{align}
respectively,
and comparing the definitions of the free-energy \cref{eq:free_energy} and \cref{eq:free_energy_thetapm}, we find the following scalings:
\begin{align}
	E_\perp^K(\kperp) \sim \frac{\taubar^2}{1+\taubar} (\kperp \rhos)^2 E_\perp^\pm(\kperp) \propto \left\{
	\begin{array}{cc}
		\displaystyle \kperp^{-5/3} , & \kperp \rhoi \ll 1, \\ [4mm]
		\displaystyle \kperp^{-1/3}, &  \kperp \rhoi \gg 1,
	\end{array}	 
	\right.
	\label{eq:kinetic_spectrum}
\end{align}
and 
\begin{align}
	E_\perp^B(\kperp) \sim \frac{1}{1 + \kperp^2 d_e^2} E_\perp^\pm(\kperp) \propto \left\{
	\begin{array}{cc}
		\displaystyle \kperp^{-5/3} , & \kperp \ll \rhoi^{-1}, d_e^{-1}, \\ [4mm]
		\displaystyle \kperp^{-7/3} , & \rhoi^{-1} \ll \kperp \ll d_e^{-1}, \\ [4mm]
		\displaystyle \kperp^{-11/3} , & d_e^{-1} \ll \kperp \ll \rhoi^{-1}, \\ [4mm]
		\displaystyle \kperp^{-13/3}, &  \kperp \gg \rhoi^{-1}, d_e^{-1}.
	\end{array}	 
	\right.
	\label{eq:magnetic_spectrum}
\end{align}

To test these predictions, we consider the simulations in \cref{tab:simulation_parameters} labelled ``high resolution''. The first two simulations, CF-res (``constant flux'') and HB-res (``helicity barrier''), differ only in their values of the electron inertial length (or, equivalently, the electron beta), having $d_e = \rhoi$ and $d_e =\rhoi/2$, respectively. The third is a simulation of RMHD, included for comparison. As we explain below, both simulations CF-res and RMHD are expected to saturate via a constant-flux cascade: this is what we indeed find, with their spectra, plotted in \cref{fig:refined_spectra}, showing good agreement with the theoretical predictions \cref{eq:thetapm_scaling}, \cref{eq:kinetic_spectrum}, and \cref{eq:magnetic_spectrum}, up to the aforementioned corrections due to intermittency. This serves as numerical confirmation of the turbulence phenomenology presented above across the RMHD, ERMHD, ultra-low-beta, and sub-$d_e$ regimes \citep[see, e.g.,][]{sch09,meyrand10,loureiro18}, at least without detailed consideration of the difficulties relating to imbalance discussed above. We note, however, that \cref{fig:refined_spectra}(a) appears to support the idea that both $\thetapm$ fields have the same scaling despite their difference in amplitude, as in \cref{eq:balanced_assumptions}, although with the caveat that they do not have the same dissipation scale. 

\subsection{Effect of helicity conservation}
\label{sec:effect_of_helicity_conservation}
As we noted in \cref{sec:nonlinear_invariants}, the fact that the generalised helicity \cref{eq:helicity} must be conserved by the system places an additional constraint on the dynamics that we did not account for in the theory presented in the preceding section. Let us rectify this now. It will be instructive to consider, for the moment, a theory of constant-flux turbulence in $\phi$ and $\Apar$ variables, rather than the $\thetapm$ ones more appropriate for imbalanced turbulence that were used in \cref{sec:constant_flux_cascade}. 

To begin, we can, from the definition \cref{eq:free_energy}, estimate the energy flux as:
\begin{align}
	\frac{1}{n_{0e} T_{0e}} \frac{\rmd W}{\rmd t} \sim  \tnl^{-1} \left(\frac{1+ \taubar}{\taubar^2}\right) \left(\frac{e{\phi}_{\kperp}}{T_{0e}}\right)^2 \sim \fluxe = \text{const},
	\label{eq:fluxe_definition}
\end{align}
where $\tnl$ is some nonlinear time associated with the cascade; we will remain agnostic to exactly what this is. Note that in writing \cref{eq:fluxe_definition}, we have assumed that the energy can be adequately represented by the first two terms in \cref{eq:free_energy}, viz., those associated with the electrostatic potential fluctuations. This is justified by the fact that the contributions to the free energy from $\phi$ and $\Apar$ must at least be in equipartition in order for the turbulence to be imbalanced [recall \cref{eq:thetapm} and \cref{eq:free_energy_thetapm}]. Analogously to \cref{eq:fluxe_definition}, we can estimate the helicity flux from \cref{eq:helicity} as 
\begin{align}
	\frac{1}{n_{0e} T_{0e}} \frac{\rmd H}{\rmd t} \sim \tnl^{-1} \frac{v_A}{c_s} \left(\frac{1 + \kperp^2 d_e^2}{\taubar}\right)  \left(\frac{e\phi_{\kperp}}{T_{0e}}\right) \left(\frac{A_{\parallel \kperp}}{\rhos B_0}\right) \cos\phase \sim \fluxh = \text{const},
	\label{eq:fluxh_definition}
\end{align}
where $\cos\phase$ is (the cosine of) some perpendicular-wavenumber-dependant phase angle between the $\phi$ and $\Apar$ fluctuations. This could be formally defined in terms of the Fourier components $\phik$ and $\Apark$ as
\begin{align}
	\cos \phase = \frac{\displaystyle\int_{-\infty}^\infty \rmd \kpar \: \Re\left< \phik (\Apark)^*\right>}{\displaystyle \left(\int_{-\infty}^\infty \rmd \kpar \: \left< | \phik|^2\right>\right)^{1/2} \left(\int_{-\infty}^\infty \rmd \kpar \: \left<|\Apark|^2 \right>\right)^{1/2}},
	\label{eq:phase_angle}
\end{align}
wherein ``$*$'' denotes the complex conjugate, and the brackets denote an ensemble average, as previously. Taking the ratio of \cref{eq:fluxh_definition} to \cref{eq:fluxe_definition}, using equipartition between the energies to relate the amplitudes of $\phi$ and $\Apar$, and recalling the definition of the injection imbalance $\sigma_\varepsilon = |\fluxh|/\fluxe$, it is straightforward to show that
\begin{align}
	\fluxh \sim  \fluxe \left( \frac{\vphase}{v_A} \right)^{-1} \cos \phase  \quad \Rightarrow \quad \cos\phase  \sim \forcing \frac{\vphase}{v_A}.
	\label{eq:phase_angle_estimate}
\end{align}
Finally, given that $0 < | \cos\phase | \leqslant 1$, it follows directly from \cref{eq:phase_angle_estimate} that
\begin{align}
	\forcing \frac{\vphase(\kperp)}{v_A}\lesssim 1.
	\label{eq:sigma_law}
\end{align}
This inequality must be satisfied everywhere in perpendicular-wavenumber space in order for the system to support a constant flux of helicity from large to small scales; if it is anywhere violated, then the assumption of constant flux breaks down. It is important to clarify that this latter statement does not only apply to the helicity flux, but also that of the free energy: if the system is not able to simultaneously cascade both invariants via constant flux, it is unable to support a constant flux of either free energy or helicity individually. This means that the nature of the turbulence is fundamentally different depending on whether or not the inequality \cref{eq:sigma_law} is satisfied; if it is, we obtain the constant-flux type turbulence discussed in \cref{sec:constant_flux_cascade}; if it is not, then the system will inevitably form a helicity barrier \citep{meyrand21,squire22}, the exact dynamics of which we shall discuss in detail in \cref{sec:the_helicity_barrier}. For now, the phrase ``helicity barrier'' can simply serve as a placeholder term for the turbulent state that occurs in the absence of a constant-flux cascade within isothermal KREHM. 

Due to the perpendicular-wavenumber dependence of the phase velocity \cref{eq:vphase}, there are a number of regimes of \cref{eq:sigma_law} to consider:
\begin{enumerate}
	\def\itemspacing{\vspace{0.2cm}}
	\itemspacing
	\item \: FLR-MHD ($\beta_e \gg m_e/m_i$) --- In this regime, the phase velocity is a strictly increasing function of perpendicular wavenumber [this is the solid black in \cref{fig:vphase}; see also \cref{eq:vphase_scalings}], meaning that, no matter the injection imbalance, \cref{eq:sigma_law} will always be violated at sufficiently small scales. The physical reason for this is clear from the first expression in \cref{eq:phase_angle_estimate}. Given a constant energy flux $\fluxe$, the increase in the phase velocity causes the helicity flux $\fluxh$ to decrease with increasing $\kperp$, which cannot be compensated for by, e.g., increasing $\cos \phase $ (i.e., further aligning the fluctuations), since the latter is bounded from above. This means that the assumption of constant flux cannot be satisfied, which, in this system, manifests itself through the formation of a helicity barrier. In a finite simulation domain, however, the resolution may not be sufficient to allow the phase velocity to increase to a value at which it would violate \cref{eq:sigma_law}. As such, if the dissipation scale of the turbulence $\kperp^\text{diss}$ is less than some critical perpendicular wavenumber 
	\begin{align}
		\kperp^\text{diss} \lesssim \kperp^\text{crit}, \quad \kperp^\text{crit} \rhoi = \frac{1}{\forcing} \left(\frac{2}{1+Z/\tau}\right)^{1/2},
		\label{eq:sigma_law_kperp_crit}
	\end{align}
	then the system is able to support a constant-flux cascade. This prediction is tested numerically in \cref{sec:breaking_the_helicity_barrier}. 
	
	\itemspacing
	\item \: Isothermal KREHM ($\beta_e \gtrsim m_e/m_i$) --- In this intermediate regime, the phase velocity increases with perpendicular wavenumber, but eventually reaches a constant value at small scales (see \cref{fig:vphase}). Using the fact that the maximum value of the phase velocity is given by the second expression in \cref{eq:vphase_limits}, we can rewrite \cref{eq:sigma_law} as 
	\begin{align}
		\beta_e \lesssim \betacrit, \quad \betacrit = \frac{2Z}{1+\tau/Z} \frac{m_e}{m_i} \frac{1}{\forcing^2}. 
		\label{eq:sigma_law_beta}
	\end{align}
	There is thus a critical value of the plasma beta below which a constant-flux cascade can occur, allowing the free energy and helicity to reach the smallest scales. In particular, \cref{eq:sigma_law_beta} predicts a constant-flux solution is always possible for $\beta_e \lesssim m_e/m_i$: it is below this value of $\beta_e$ that the phase velocity begins to decrease with perpendicular wavenumber, and we find ourselves in the ultra-low beta regime. 
	
	\itemspacing
	\item \: Ultra-low beta ($\beta_e \ll m_e/m_i$) --- This is the opposite of regime (i), in that the phase velocity is a strictly decreasing function of perpendicular wavenumber [see \cref{eq:vphase_scalings}]. This means that, since $\forcing \leqslant 1$, the inequality \cref{eq:sigma_law} will always be satisfied, regardless of the injection imbalance. This once again follows from the first expression in \cref{eq:phase_angle_estimate}: the decrease in the phase velocity would cause $\fluxh$ to increase with increasing $\kperp$, but, unlike in regime (i), this always can be compensated for by decreasing $\cos \phase$, and so a constant-flux cascade is always allowed. This is the phenomenon of ``dynamic phase alignment'' \citep{loureiro18,milanese20}, wherein $\phi$ and $\Apar$ fluctuations become increasingly misaligned at small scales in order to maintain a constant flux of both free energy and helicity. The relationship between the helicity barrier and dynamic phase alignment is discussed further in \cref{sec:dynamic_phase_alignment}. 
		
	\itemspacing
\end{enumerate}
It is worth noting that while the inequality \cref{eq:sigma_law} was derived here using the above heuristic scaling arguments, it can be put on a rigorous footing. In regimes (i) and (ii), where the phase velocity \cref{eq:vphase} is an increasing function of perpendicular wavenumber, it is possible to prove --- following an argument identical to that used in \cite{meyrand21} \citep[see also][]{alexakis18} --- that the perpendicular fluxes of free energy $\Pi_W(\kperp)$ and helicity $\Pi_H(\kperp)$ must satisfy the inequality $|\Pi_H(\kperp)|/|\Pi_W(\kperp)| \leqslant v_A/\vphase(\kperp)$ in order for a constant-flux solution to exist [the presence of the phase velocity arises from the perpendicular-wavenumber dependence of the scale-by-scale helicity; see \cref{eq:helicity_thetapm}]. This inequality implies \cref{eq:sigma_law}, since $|\Pi_H(\kperp)|/|\Pi_W(\kperp)| = \forcing$ in a constant-flux cascade, by definition. One cannot apply the same argument in regime (iii), where the phase velocity is a decreasing function of perpendicular wavenumber --- this is inconsequential, however, because \cref{eq:sigma_law} will always be satisfied in this regime, and so a constant-flux solution is always possible. Given, then, that the inequality \cref{eq:sigma_law} can only be violated in the regimes where rigorous proof exists, we regard it as a stronger condition than might be implied by the heuristic arguments that were initially used to derive it. The resultant prediction, and subsequent numerical confirmation, of the existence of the critical beta \cref{eq:sigma_law_beta} is the central result of this paper. However, in order to be able to test this prediction, one must have an understanding of the dynamics associated with the helicity barrier, to which we now turn our focus. The impatient reader, or one already familiar with the helicity barrier, may wish to skip ahead to \cref{sec:breaking_the_helicity_barrier}, working backwards where further clarification is required. 

\subsection{The helicity barrier}
\label{sec:the_helicity_barrier}
To explore the dynamics of the helicity barrier within isothermal KREHM, we compare simulations CF and HB from the set ``comparison'' in \cref{tab:simulation_parameters}. Importantly, these simulations lie on either side of the critical beta \cref{eq:sigma_law_beta}, with values of $d_e$ that differ by a factor of two, but have parameters that are otherwise identical. The injection imbalance of $\forcing = 0.8$ used for both simulations corresponds to a critical beta $\betacrit (m_e/m_i)^{-1} = 1.56$ --- simulation CF, which saturates via constant flux, has $\beta_e / \betacrit = 0.64$, while simulation HB, which forms a helicity barrier, has $\beta_e/\betacrit = 2.56$. The choice of these specific values, however, is inconsequential: simulations on either side of $\betacrit$ display almost identical behaviour regardless of the choice of $\forcing$. We have chosen these particular simulations because they provide illustrative examples of the behaviour of the system with and without the helicity barrier, despite their very similar parameters.

\subsubsection{Energy fluxes}
\label{sec:energy_fluxes}
\begin{figure}
	
	\centering
	
	\begin{tabular}{c}
		
		\includegraphics[width=1\textwidth]{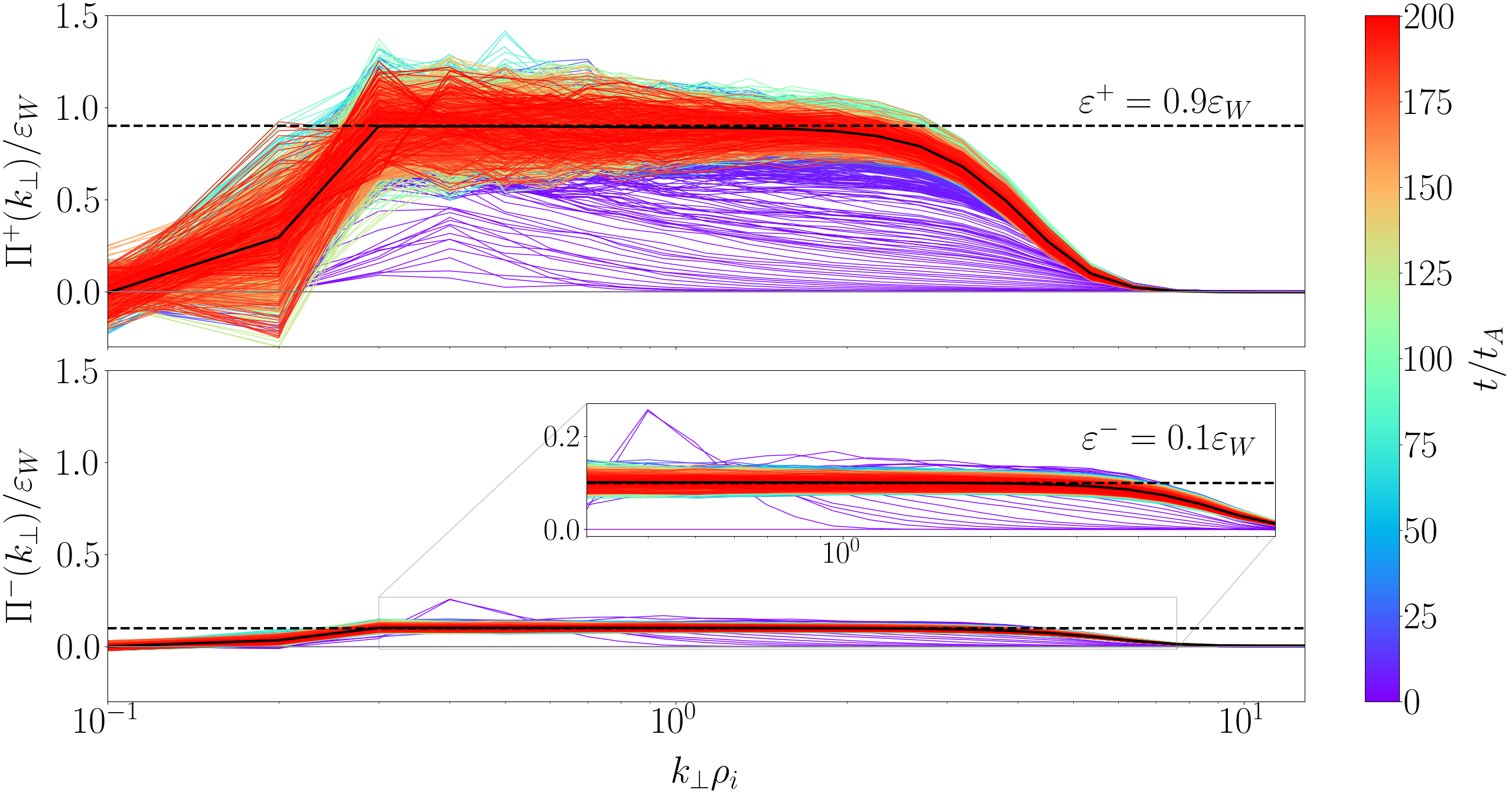}  \\\\
		(a) Simulation CF (constant flux) \\\\
		\includegraphics[width=1\textwidth]{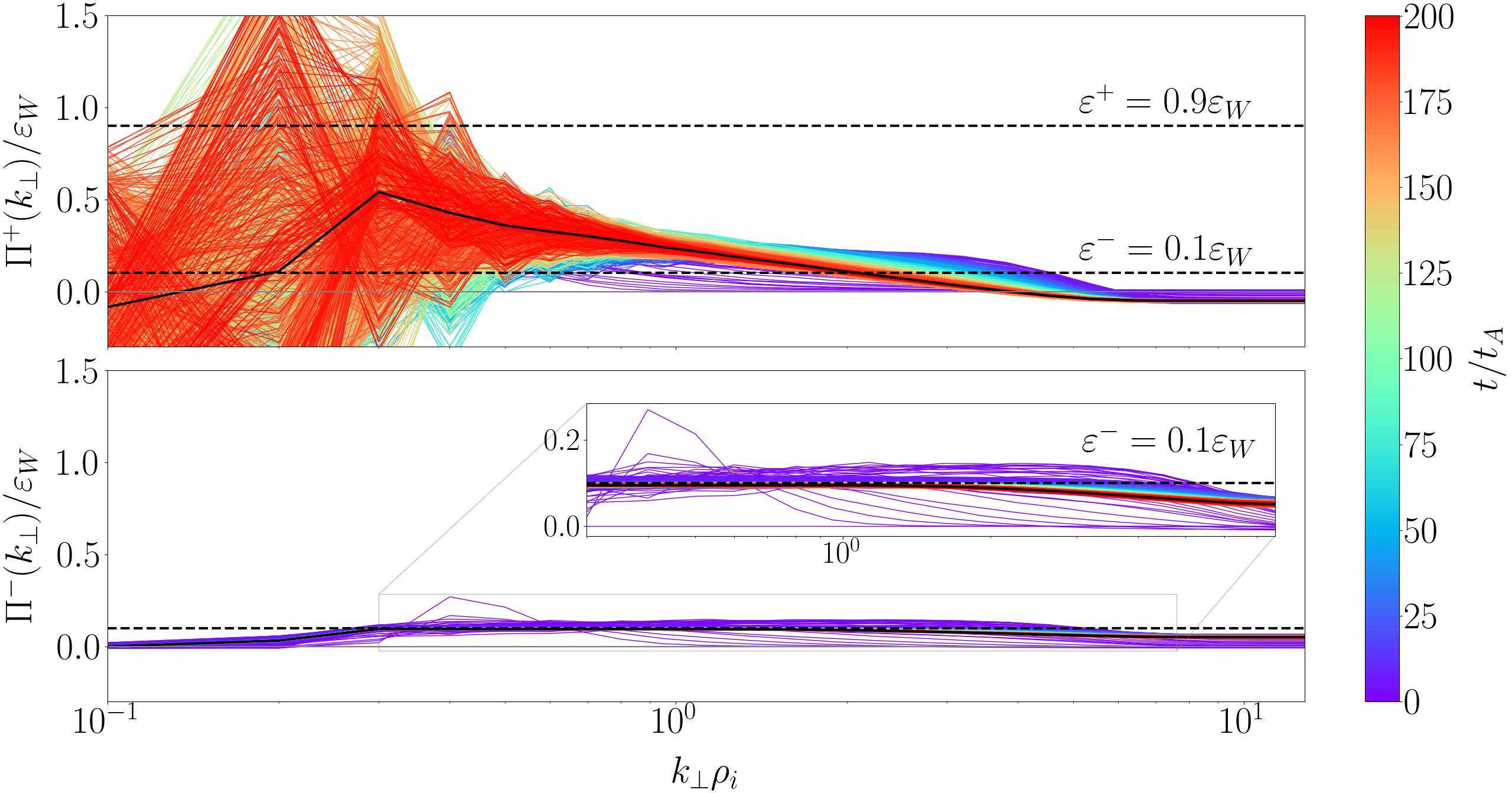}  \\\\
		(b) Simulation HB (helicity barrier)
		
	\end{tabular}
	
	\caption[]{Time-evolution of the spectral energy fluxes $\Pi^\pm(\kperp)$ computed directly from the nonlinear terms in \cref{eq:phi_equation} and \cref{eq:apar_equation}, normalised to the the total energy flux $\fluxe$. Simulations CF and HB from \cref{tab:simulation_parameters} are shown in panels (a) and (b), respectively. The colours indicate the value of time corresponding to a given line, while the solid black lines correspond to the average value of each flux over the last $20\%$ of the simulation time. The horizontal dashed lines indicate the values of the flux \cref{eq:fluxpm} expected if the system is able to maintain a constant flux; we have included a line corresponding to $\fluxm$ in the upper panel of (b) for ease of comparison. It is clear that the total flux reaching small scales in the presence of the helicity barrier is significantly smaller than in the constant flux case (note that, in both cases, the decrease in the flux at small scales is due to the presence of perpendicular hyperdissipation).}
	\label{fig:fluxes}
\end{figure}

In \cref{fig:fluxes}, we plot the spectral energy fluxes $\Pi^\pm(\kperp)$ associated with the potentials $\thetapm$ from these two simulations, calculated by summing the contributions of the nonlinear transfers above and below the particular $\kperp$ of interest. Simulation CF, with $\beta_e/\betacrit < 1$, behaves as expected: the fluxes of both $\thetap$ and $\thetam$ are approximately stationary (in time), constant (as a function of perpendicular wavenumber), and equal to their injected values \cref{eq:fluxpm}.
Note, from \cref{fig:fluxes}(a), that the entirety of the injected free energy $\fluxe = \fluxp + \fluxm$ is carried by the turbulence to small scales, where it is then dissipated by the perpendicular hyperdissipation \cref{eq:hyperdissipation}. This provides \textit{a posteriori} justification of the constant-flux assumption made in \cref{sec:constant_flux_cascade}. 

The simulation HB, with $\beta_e /\betacrit > 1$, however, displays very different behaviour. While the flux of $\thetam$ remains approximately stationary and constant at $\fluxm$ (more so, in fact, than simulation CF), the flux of $\thetap$ in the inertial range is a decreasing function of both time and perpendicular wavenumber, while the largest scales display rapid fluctuations. Of particular note, readily apparent from \cref{fig:fluxes}(b), is the fact that only a small fraction of the injected energy is cascaded to small scales, remaining limited to $\approx 2\fluxm = (1-\forcing)\fluxe$, irrespective of the turbulent amplitudes, which are much larger than in simulation CF (see \cref{fig:spectra_time_evolution}). This is the helicity barrier: the breakdown of the constant-flux assumption, due to a violation of the inequality \cref{eq:sigma_law}, causes the system to form a ``barrier'' that prevents all but the balanced portion of the injected energy from cascading to small perpendicular scales.

\subsubsection{Dissipation}
\label{sec:dissipation}
The remaining energy must, therefore, find another route to thermalisation, which it does so by accessing small \textit{parallel} scales. In \cref{fig:dissipation_rates}, we plot the free energy $W$ and its parallel and perpendicular (hyper-)dissipation rates, denoted $\dpara$ and $\dperp$, respectively, as a function of time for both simulations. It is clear that $\dpara \ll \dperp$ at all times in simulation CF, allowing the free energy to quickly saturate on perpendicular dissipation, as expected from a system undergoing a constant-flux cascade. Conversely, in simulation HB, the ratio of the parallel dissipation rate to the total dissipation rate, viz.,  
\begin{align}
	\dissratio = \frac{\dpara}{\dpara + \dperp},
	\label{eq:dissipation_ratio}
\end{align}
is an increasing function of time; we plot this explicitly in the lower panel of \cref{fig:dissipation_rates}.
Both this ratio (henceforth termed the \textit{dissipation ratio}) and the free energy eventually saturate at late times when the large-scale turbulent amplitudes reach sufficiently high levels that the energy removed by parallel dissipation (at small parallel scales) can balance the fraction of the injected energy that is unable to cascade \citep{meyrand21}. This saturation is, of course, unphysical, since it breaks the assumption of anisotropy ($\kpar \ll \kperp$) used to derive the isothermal KREHM system, and depends on the details of the parallel dissipation (e.g., the specific value of $\nu_z$; see \citealt{meyrand21}). As such, the only dynamics relevant to real physical systems are those that occur before this saturation, a period of time that we shall henceforth refer to as the \textit{pseudostationary phase}.

\begin{figure}
	
	\includegraphics[width=1\textwidth]{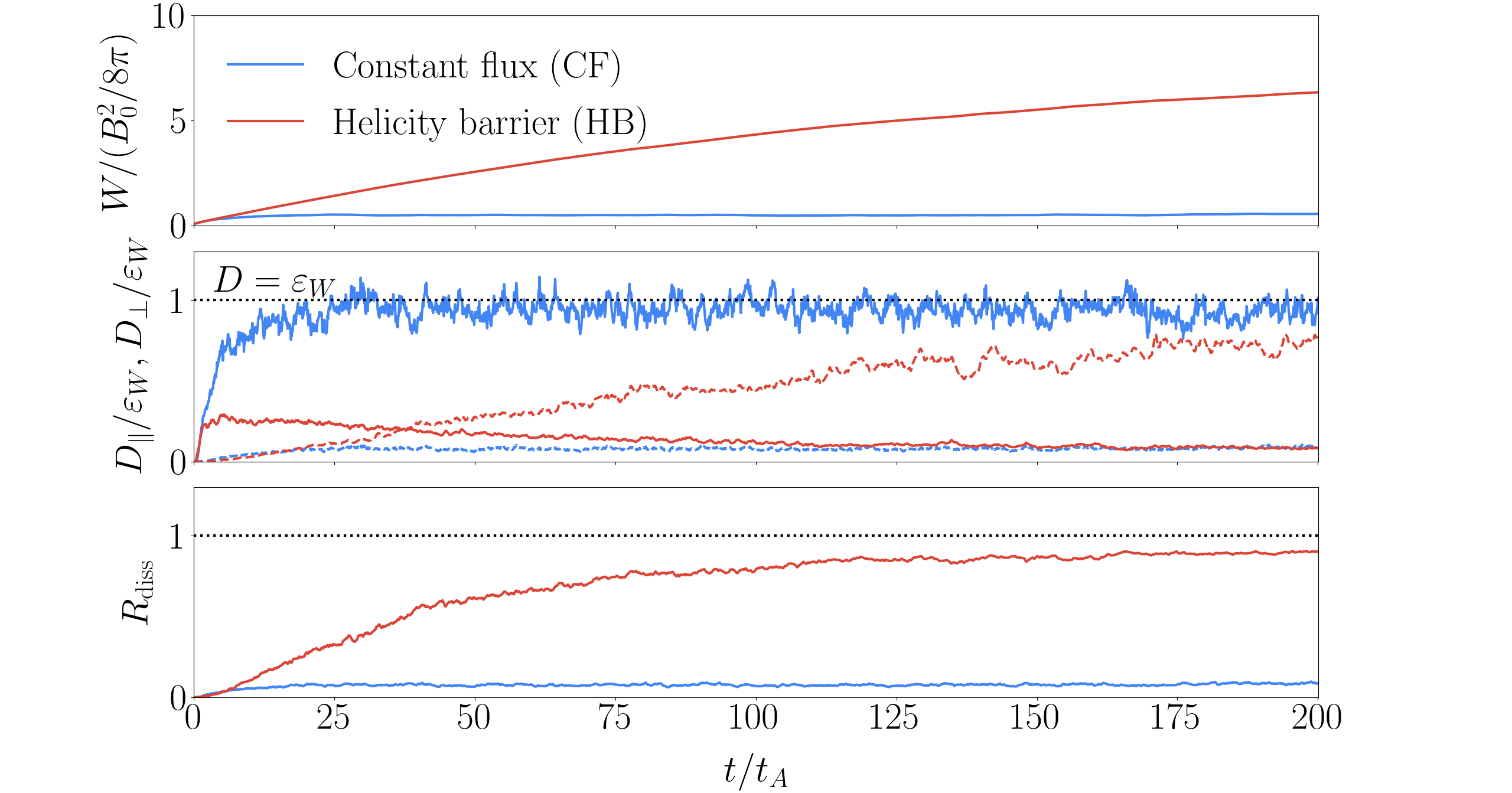}
	
	\caption[]{The free energy and its dissipation rates as a function of time for simulations CF (blue) and HB (red). The top panel shows the free energy, while the middle panel shows the parallel and perpendicular dissipation rates, $\dpara$ and $\dperp$, in dashed and solid lines, respectively. The bottom panel shows the dissipation ratio \cref{eq:dissipation_ratio}. It is clear that $\dissratio \ll 1$ for~the constant-flux case, while $\dissratio$ grows slowly to $\approx 1$ for the helicity-barrier one.}
	\label{fig:dissipation_rates}
\end{figure} 

\subsubsection{Perpendicular energy spectra}
\label{sec:perpendicular_energy_spectra}
\begin{figure}
	
	\centering
	
	\begin{tabular}{c}
		
		\includegraphics[width=1\textwidth]{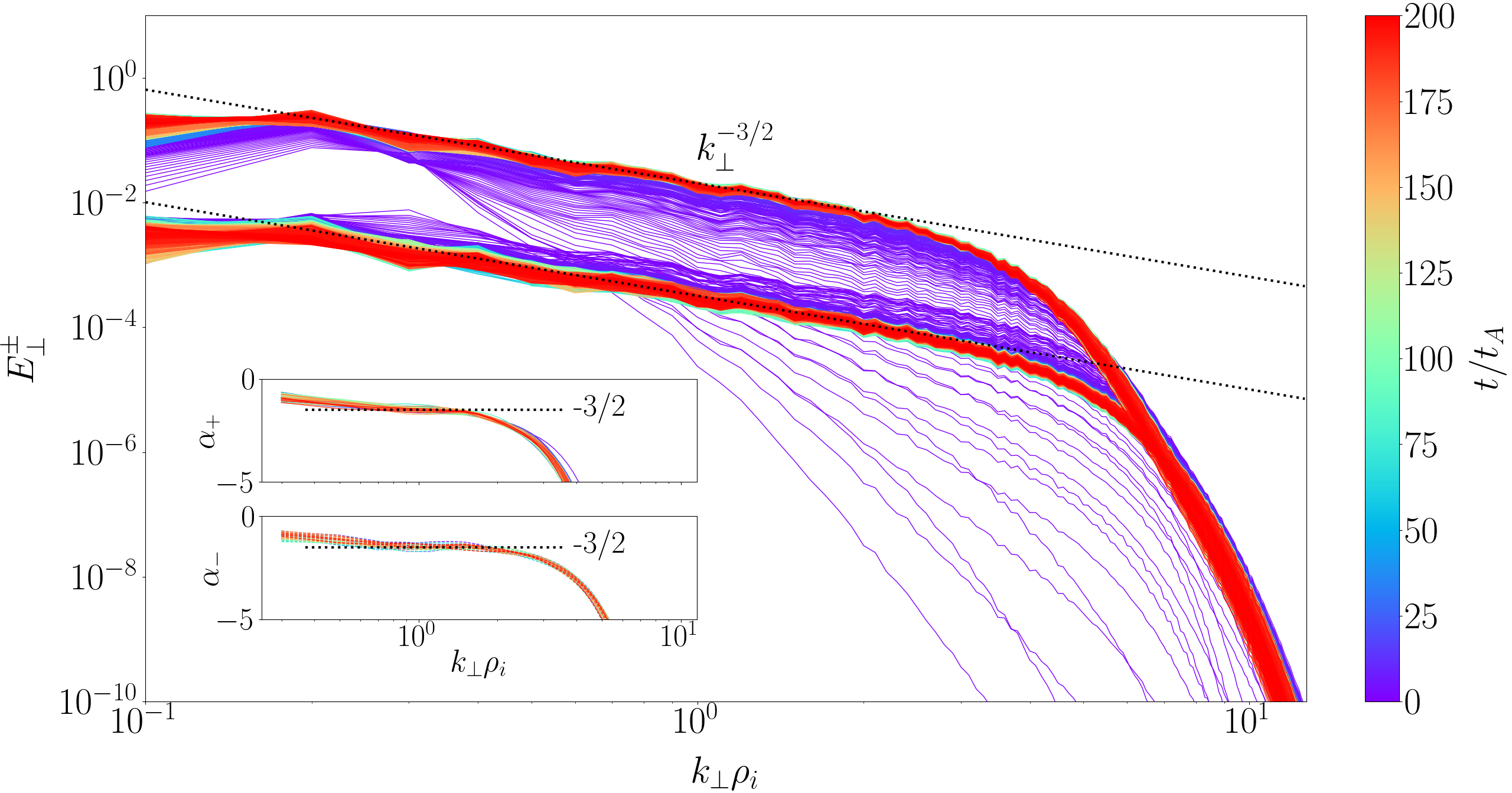}  \\\\
		(a) Simulation CF (constant flux) \\\\
		\includegraphics[width=1\textwidth]{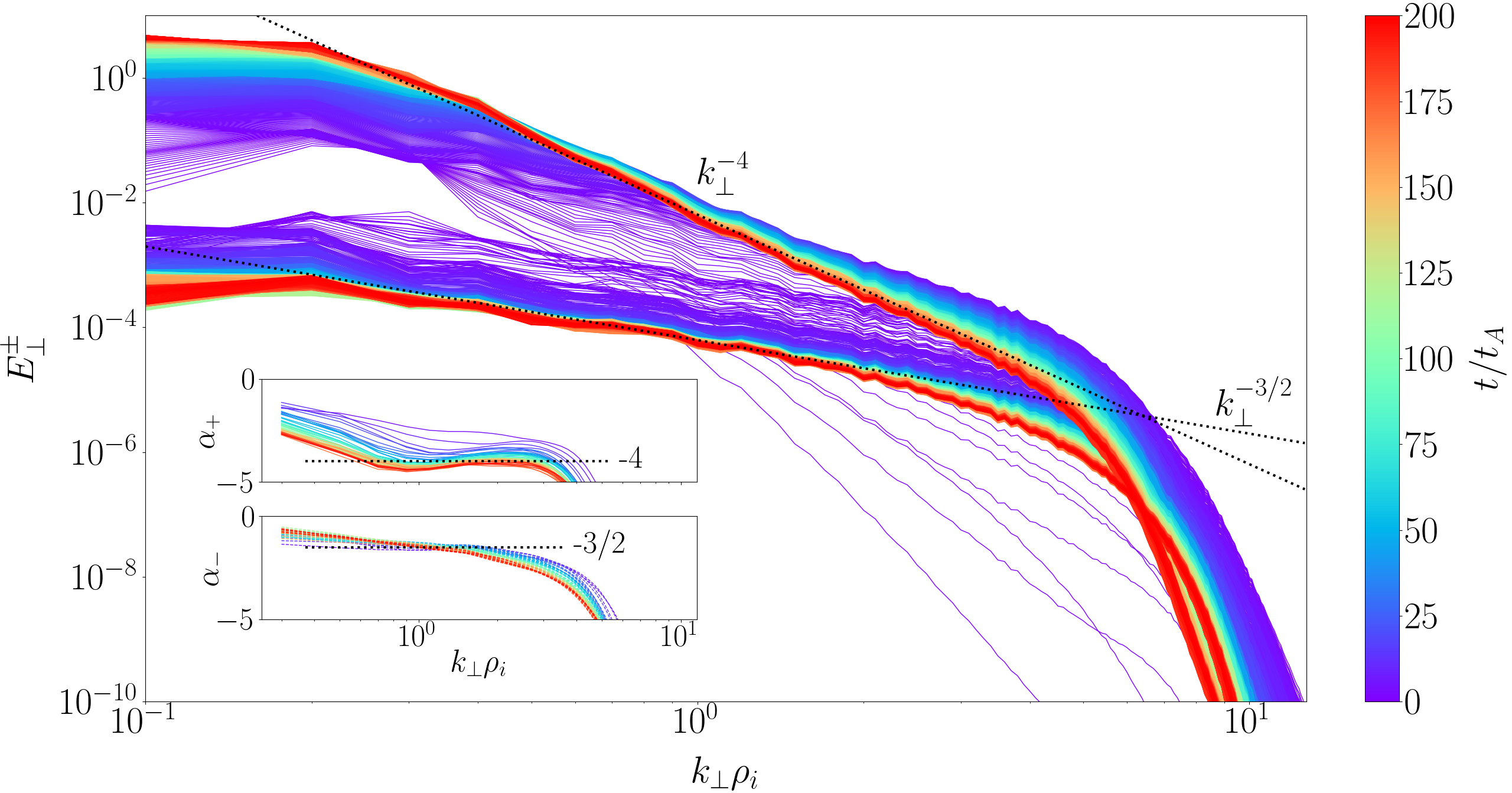}  \\\\
		(b) Simulation HB (helicity barrier)
		
	\end{tabular}
	
	\caption[]{Time evolution of the perpendicular spectra $E_\perp^\pm(\kperp)$ for simulations CF and HB in \cref{tab:simulation_parameters}, shown in panels (a) and (b), respectively. The colours indicate the value of time corresponding to a given line, while the dotted black lines indicate approximate scalings of the spectra. The inset panels show the scaling exponents $\alpha^\pm = \rmd \log E_\perp^\pm/\rmd \log \kperp$ for each spectrum. In the helicity barrier case, the $E_\perp^+(\kperp)$ spectrum clearly forms a spectral break that moves towards large scales in time.}
	\label{fig:spectra_time_evolution}
\end{figure}

What is the impact of these dynamics on the measured perpendicular energy spectra? In \cref{fig:spectra_time_evolution}, we plot the time evolution of the $E_\perp^\pm(\kperp)$ spectra \cref{eq:thetapm_spectra_def} for both simulations, with their high-resolution counterparts (see \cref{tab:simulation_parameters}) plotted in \cref{fig:refined_spectra}(a). As expected, simulation CF agrees well with the constant-flux cascade predictions of \cref{sec:constant_flux_cascade}: both $E_\perp^\pm(\kperp)$ have approximately the same $\sim \kperp^{-3/2}$ slope in the inertial range, differing by only their outer-scale amplitudes. In simulation HB, the weaker spectrum $E_\perp^-(\kperp)$ behaves similarly, quickly saturating with a $\sim \kperp^{-3/2}$ slope. However, given that the $\thetap$ flux is a decreasing function of perpendicular wavenumber, the associated $E_\perp^+(\kperp)$ spectrum cannot reach such a stationary state. Instead, during the start of the pseudostationary phase, it forms a spectral break around $\kperp \rhoi \sim 1$, with the location of this spectral break then migrating to larger scales. The slope below the break is consistently measured to be $\sim \kperp^{-4}$, across all simulations in \cref{tab:simulation_parameters} that exhibited a helicity barrier. The higher resolution simulation HB-res (see \cref{tab:simulation_parameters}) plotted in \cref{fig:refined_spectra} also shows a spectral flattening at smaller scales $\kperp \rhoi \gtrsim 1$, consistent with observations of the spectral ``transition range'' in near-sun solar wind plasmas \citep{bowen20,duan21,bowen24}. Additionally, the reduction of the $\thetap$ flux to small scales (in comparison to simulation CF; see \cref{fig:fluxes}) causes the outer-scale energy, and thus the normalised cross-helicity $\sigma_c$, to increase in time (while both the energy injection rate $\fluxe$ and the injection imbalance $\forcing$ remain constant). As discussed above, this will continue until the large-scale amplitudes reach sufficient levels that saturation can occur on parallel dissipation. Both \cite{meyrand21} and \cite{squire23} found that the position of the break is correlated with $\sigma_c$ finding that, approximately, its location in perpendicular-wavenumber space evolved as $\kperp \rhoi \sim (1-\sigma_c)^{1/4}$. We have not attempted to verify such a scaling here due to the relatively small inertial range at this resolution, and the fact that said scaling may be complicated by the finite $d_e$ effects present in these simulations. Nevertheless, while a dynamical theory that explains these features remains the focus of ongoing work, the existence of the break, as well as the steep spectral scaling below it, are both persistent features of helicity-barrier-mediated turbulence.

\begin{figure}
	
	\includegraphics[width=1\textwidth]{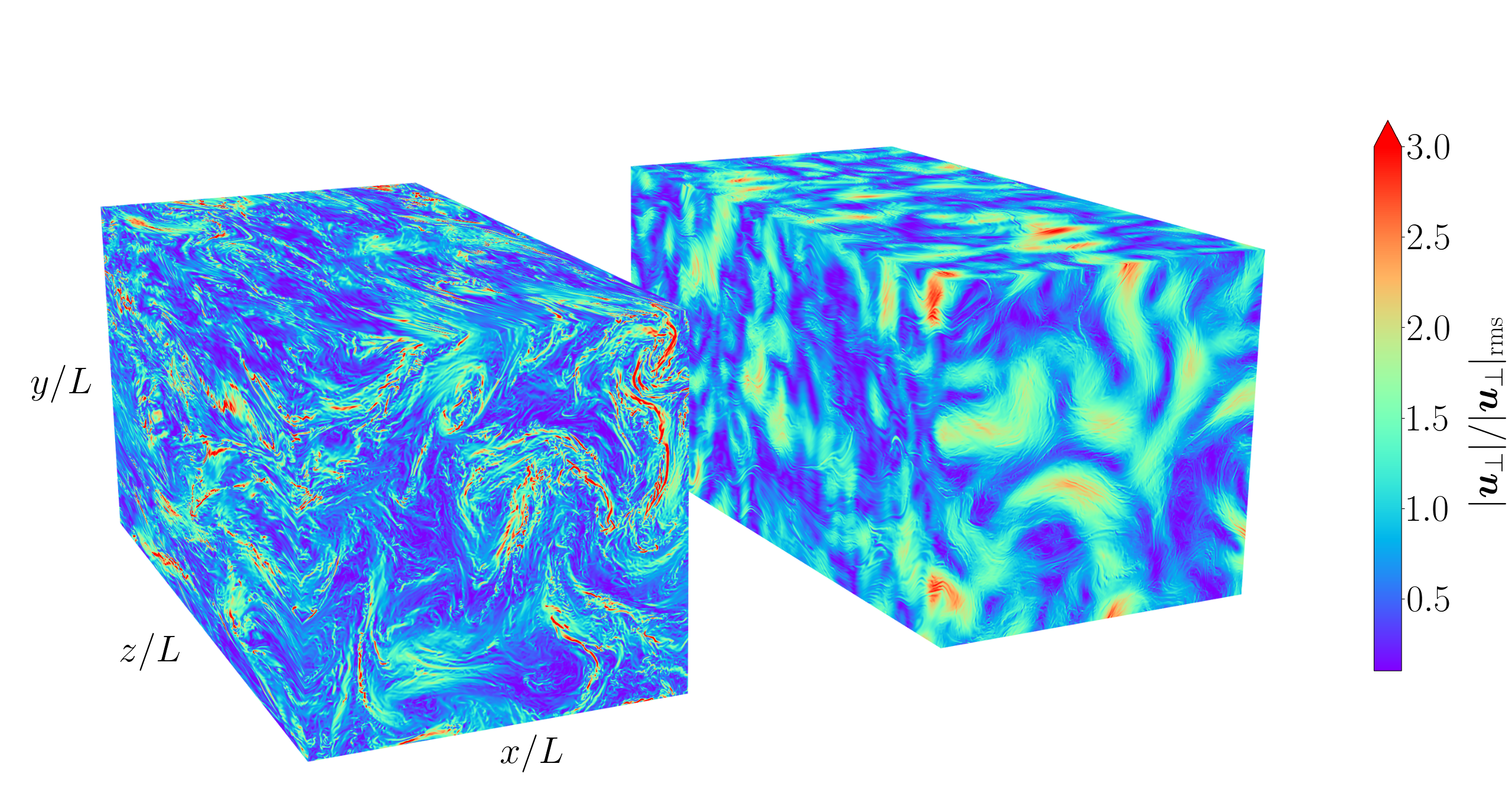}
	
	\caption[]{Real-space snapshots of the $\vec{E} \times \vec{B}$ flow $\uperp$ [see \cref{eq:rmdt}] for the simulations CF-res (left) and HB-res (right). The colours indicate the magnitude of $\uperp$ relative to its (spatial) root-mean-square value, while the coordinate directions are as shown. While the structure of the turbulence in simulation CF-res is typical of a constant-flux cascade [though note the significant small-scale plasmoid activity due to finite $d_e$ effects; see \cite{zhou23MNRAS}], this is not the case for simulation HB-res: the majority of the energy resides in large-scale structures because it is prevented from cascading to small scales by the helicity barrier. The dramatic difference between these two cases is made even more surprising by the fact that the simulations differ only in their value of the electron inertial length, having $d_e = \rhoi$ and $d_e = \rhoi/2$, respectively.}
	\label{fig:realspace_data}
\end{figure} 

It should be clear from the above discussion that the helicity barrier state is dramatically different from that associated with a constant-flux cascade, viz., the nature of the turbulence is fundamentally changed depending on whether or not the inequality \cref{eq:sigma_law} is satisfied. The surprising element of this is the fact that these two states can lie so close to one another in parameter space: we recall that the electron inertial scales for the simulations that we have been considering in this section only differ by a factor of two, being $d_e = \rhoi$ for the constant-flux case, and $d_e = \rhoi/2$ for the helicity-barrier one. Indeed, the real-space snapshots of the turbulence shown in \cref{fig:realspace_data} are completely different, despite this small difference in physical parameters.

\subsection{Breaking the barrier}
\label{sec:breaking_the_helicity_barrier}
To briefly summarise the findings of the previous section, the helicity barrier has three key features: (i) it only allows the balanced fraction ($\approx 2\fluxm$) of the free energy to cascade to small scales (\cref{sec:energy_fluxes}); (ii) the remainder of the free energy remains at large perpendicular scales where it dissipates on small parallel ones, meaning that the ratio of the parallel dissipation to the total dissipation $\dissratio$ [see \cref{eq:dissipation_ratio}] is an increasing function of time during the pseudostationary phase (\cref{sec:dissipation}); and (iii) the spectrum of $\thetap$ displays a sharp spectral break, with an approximate $\sim \kperp^{-4}$ scaling below it, which widens (moves towards larger scales) over time (\cref{sec:perpendicular_energy_spectra}). 

We now wish to test the predictions of \cref{sec:effect_of_helicity_conservation} across a wide range of $\beta_e$ and $\forcing$. This requires the ability to efficiently determine whether or not a helicity barrier has formed in a given simulation. While feature (i) is the clearest measure of the helicity barrier, calculating the spectral energy fluxes $\Pi^\pm(\kperp)$ is computationally expensive, making it unfeasible to use across a large simulation set. Similarly, feature (iii) is not a reliable measure of helicity-barrier formation at early times because it requires the spectral slopes both above and below the break to be sufficiently well resolved, which is not always possible at lower resolutions. As such, we choose to exploit the fact that the dissipation ratio \cref{eq:dissipation_ratio} is an increasing function of time during the pseudostationary phase, viz., simulations that, on average, have $\rmd\dissratio/\rmd t \approx 0$ will be undergoing a constant-flux cascade, while those with $\rmd \dissratio/\rmd t \gtrsim 0$ will have formed a helicity barrier. This is by no means a unique measure of helicity-barrier formation, but will prove sufficient and appropriate for our purposes here. We emphasise that while the dissipation ratio provides a useful measure when applied to our simulations, it should not be viewed as a diagnostic to be measured with spacecraft data, wherein other features of the helicity barrier (such as energy fluxes, spectral slopes, or implied heating rates) would be more appropriate. Indeed, it is only a useful measure 
within subsidiary limits of gyrokinetics like isothermal KREHM, since artificial parallel dissipation must be added to allow saturation in the absence of other dissipative mechanisms on small parallel scales at $\kperp \rhoi \lesssim 1$ (e.g., ICW heating of ions around $\kpar d_e \sim 1$, which lies outside the gyrokinetic approximation).	

\begin{figure}
	
	\centering
	
	\begin{tabular}{c}
		
		\includegraphics[width=1\textwidth]{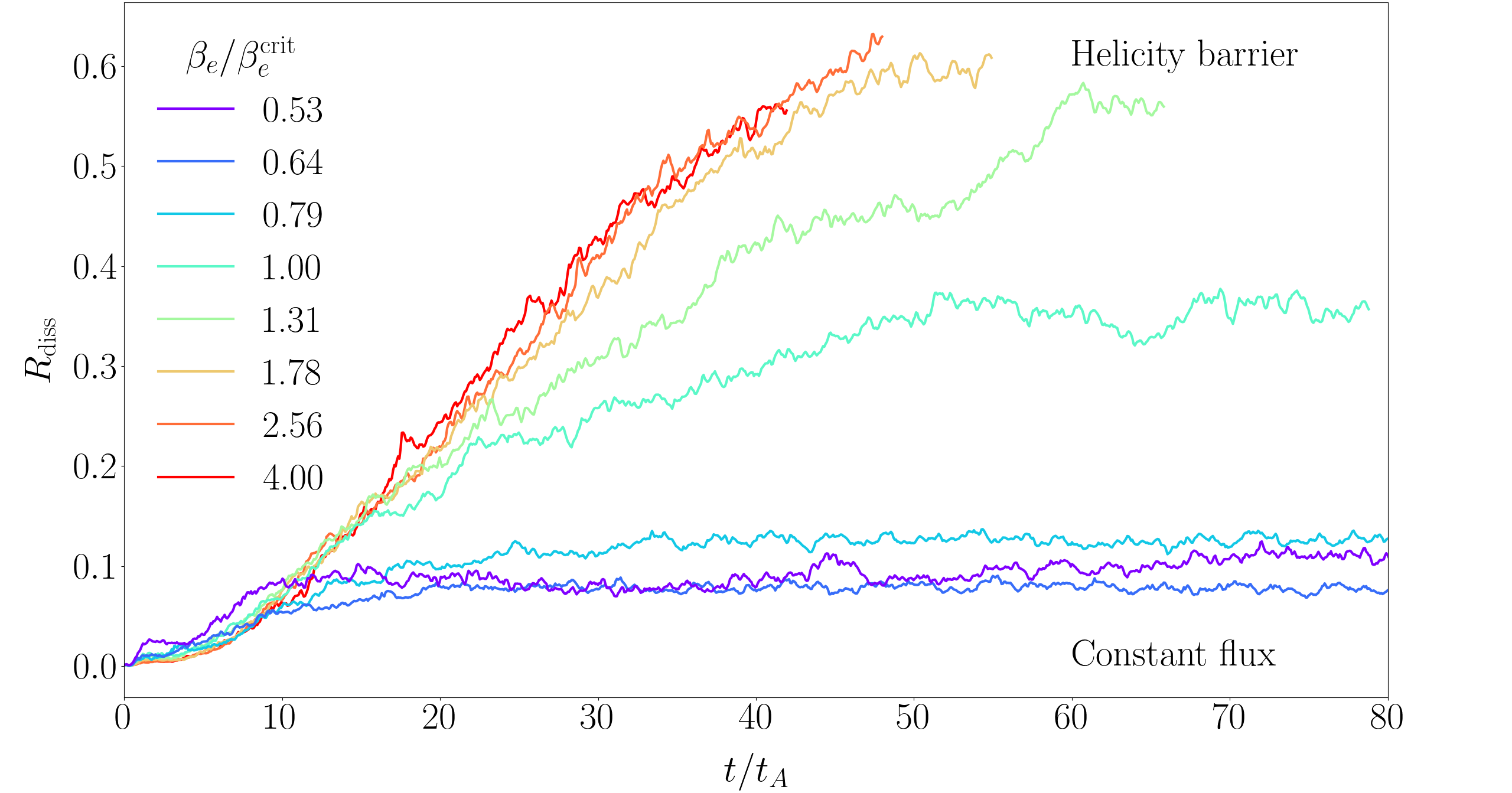}  \\\\
		(a) Dissipation ratio \cref{eq:dissipation_ratio} \\\\\\
		\includegraphics[width=1\textwidth]{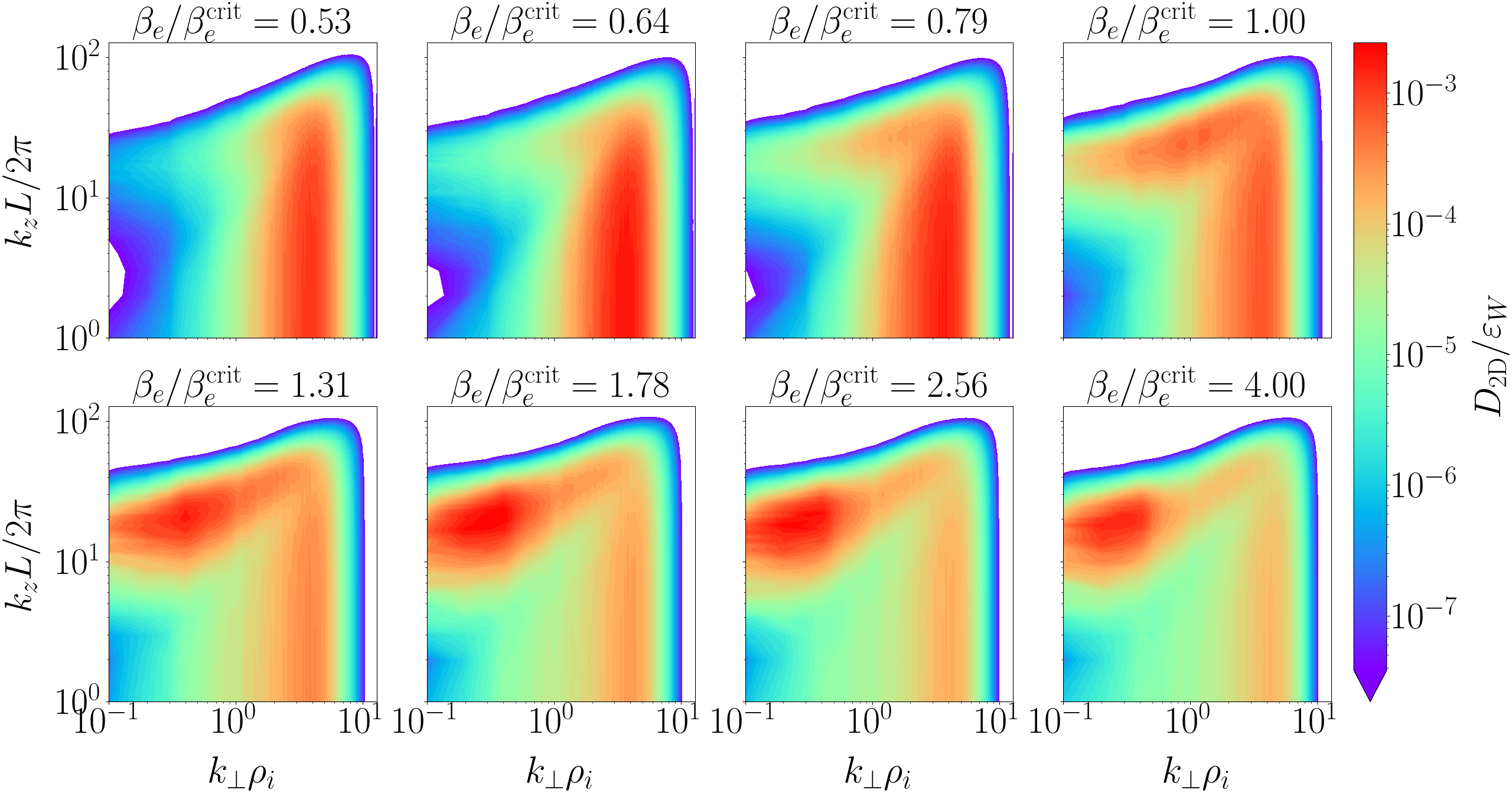}  \\\\
		(b) Two-dimensional dissipation spectrum
		
	\end{tabular}
	
	\caption[]{Measures of the dissipation associated with the set of eight simulations with $\forcing = 0.8$ from the set labelled ``beta scan'' in \cref{tab:simulation_parameters}. (a) The dissipation ratio \cref{eq:dissipation_ratio} plotted as a function of time. The colours indicate the value of $\beta/\betacrit$ for each simulation. (b) The two-dimensional spectrum of the dissipation in the $(\kperp, k_z)$ plane, averaged over the last $20\%$ of the simulation time, and normalised to the energy injection rate $\fluxe$, with amplitude as indicated by the colourbar.}
	\label{fig:dissipation_spectrum}
\end{figure} 

To explicitly demonstrate the utility of using $\dissratio$ as a diagnostic, we consider the series of simulations with $\forcing = 0.8$ from the set labelled ``beta scan'' in \cref{tab:simulation_parameters}, in which $\beta_e/\betacrit$ is varied between $0.53$ and $4.00$. In \cref{fig:dissipation_spectrum}(a), we plot $\dissratio$ as a function of time for each simulation. It is clear the simulations are split between those that are approximately steady in time (constant flux), and those that increase with time (helicity barrier). This is illustrated in more detail in \cref{fig:dissipation_spectrum}(b), in which we plot the two-dimensional spectrum of the dissipation $D_\text{2D}(\kperp, k_z)$ for the same set of simulations. Those whose dissipation ratio is approximately constant in time dissipate almost all of their energy at small perpendicular scales (right-hand side of the plot), while those whose dissipation ratio is growing have more dissipation at small parallel scales and $\kperp \rhoi < 1$ (top left of the plot), a clear signature of the helicity barrier.\footnote{We note that while the dissipation on these scales looks prominent in \cref{fig:dissipation_spectrum}(b), these simulations have not yet reached saturation, as is evident from \cref{fig:dissipation_spectrum}(a).} The behaviour of the simulation with $\beta/\betacrit = 1$, somewhat unsurprisingly, lies intermediate between these two states: the dissipation ratio initially grows in time during a short pseudostationary phase [\cref{fig:dissipation_spectrum}(a)], before achieving saturation with similar levels of parallel and perpendicular dissipation [\cref{fig:dissipation_spectrum}(b)]. 

It is clear that the rate of change of the dissipation ratio gives a clear measure of helicity barrier formation within this set of simulations. However, we require a criterion for helicity-barrier formation that will apply across the full range of imbalances considered in \cref{tab:simulation_parameters}. For this, we choose to consider the non-dimensionalised rate of change of the dissipation ratio given by $(\rmd \dissratio/\rmd t) (\fluxh/H)^{-1}$. The normalising factor $\fluxh/H$ --- the ratio of the helicity injection rate to the helicity itself --- is effectively a measure of the timescale over which the imbalanced portion of the energy [compare \cref{eq:free_energy_thetapm} and \cref{eq:helicity_thetapm}] builds up on scales $\kperp \rhoi \lesssim 1$ (recall, from \cref{sec:energy_fluxes}, that only the balanced portion $\approx 2\fluxm$ is allowed to cascade to small scales). Given that $\fluxh = \forcing \fluxe$, this normalisation accounts for both variations in the overall energy injection rate $\fluxe$, as well as the fact that systems with lower imbalances will typically have smaller $\rmd \dissratio/\rmd t$ even when a helicity barrier has formed (the lower imbalance means that it takes longer for sufficient energy to build up on the largest perpendicular scales and begin dissipating on small parallel ones). Using the above set of simulations, we find that a time-averaged value of
\begin{align}
	\left[\frac{\rmd \dissratio}{\rmd t } \left(\frac{\fluxh}{H}\right)^{-1}\right]^\text{crit} = 0.25
	\label{eq:dissipation_ratio_critical}
\end{align}
is reasonable to distinguish between the constant-flux and helicity-barrier regimes. The time average is performed over the last 80\% of the simulation time in order to exclude the initial transient phase that occurs in the constant-flux simulations.\footnote{This transient phase results from the finite time required for the free energy injected on the largest scales to cascade to the smallest ones; our simulations are initialised with low-amplitude noise at all scales, and so it takes a number of nonlinear turnover times for energy to reach the dissipation scale. This is manifest in, e.g., \cref{fig:spectra_time_evolution}(a), wherein the spectra are ``depleted'' on the largest perpendicular wavenumbers at early times (purple lines). Note that this initial transient also occurs in the simulations that form a helicity barrier [see, e.g., \cref{fig:spectra_time_evolution}(b)], though the presence of the pseudostationary phase makes the length of this initial transient less obvious in \cref{fig:dissipation_spectrum}(a).} We note that while simulations with $\beta/\betacrit \approx 1$ could be classified differently depending on the exact value chosen in \cref{eq:dissipation_ratio_critical}, the classification of simulations with $\beta/\betacrit$ very different from unity is robust to such choices. As such, we will henceforth use the numerical value \cref{eq:dissipation_ratio_critical} as our criterion for determining helicity-barrier formation, as applied to the simulations considered in \cref{tab:simulation_parameters}. Let us now test the two predictions made in \cref{sec:effect_of_helicity_conservation}: that the formation of the helicity barrier depends on: (i) having a plasma beta above the critical value \cref{eq:sigma_law_beta}; and (ii) resolving the critical perpendicular wavenumber \cref{eq:sigma_law_kperp_crit} the dissipation range. 

For the former, we consider the set of simulations labelled ``beta scan'' in \cref{tab:simulation_parameters}, in which the injection imbalance $\forcing$ is varied between $0.1$ and $0.99$, while $\beta_e (m_e/m_i)^{-1}$ is varied between $0.83$ and $625$ (corresponding to a variation in $\beta_e/\betacrit$ between $0.03$ and $25$). Note that as $\forcing$ is varied, the forcing is also modified in such a way as to keep the amplitude of the stronger field $\thetap$ approximately constant according to the first expression in \cref{eq:field_ratio}, i.e., the forcing is scaled to keep $(\fluxp)^2/\fluxm \propto (1-\forcing)/(1+\forcing)^2 = \text{const}$. This was done in order to minimise the possible effect of the outer-scale forcing on helicity barrier formation. The results of this scan are plotted in \cref{fig:beta_scan}, which displays excellent agreement with \cref{eq:sigma_law_beta} over multiple orders of magnitude in $\beta_e (m_e/m_i)^{-1}$. The inset panel demonstrates that the criterion that we have applied to determine helicity barrier formation is very well satisfied: there is a rapid increase in $(\rmd \dissratio/\rmd t) (\fluxh/H)^{-1}$ around $\beta_e/\betacrit = 1$ across all sets of simulations, consistent with this being the location in parameter space where the helicity barrier forms. We thus confirm the prediction \cref{eq:sigma_law_beta} of the critical line in $(\forcing,\beta_e)$ parameter space above which a helicity barrier will always form; the implications of this result are discussed in \cref{sec:summary_and_discussion}.

\begin{figure}
	
	\includegraphics[width=1\textwidth]{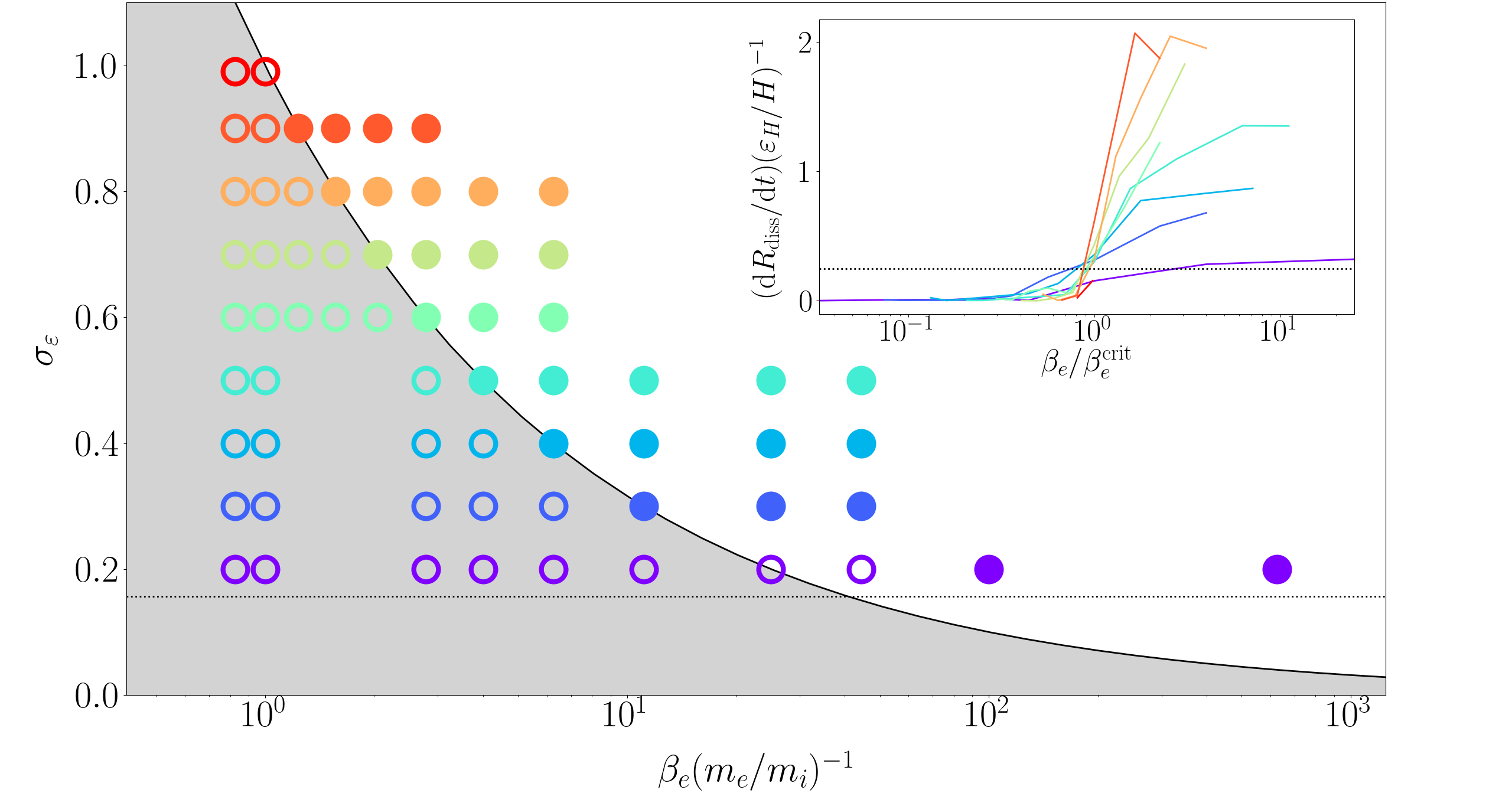}
	
	\caption[]{Data from a two-dimensional parameter scan of injection imbalance $\forcing$ versus the (normalised) electron plasma beta $\beta_e (m_e/m_i)^{-1}$. Each circle corresponds to a simulation from the set ``beta scan'' in \cref{tab:simulation_parameters}, with filled/open circles indicating the presence/absence of the helicity barrier. The solid black line corresponds to the value of $\forcing$ below which, for a given $\beta_e$, the helicity barrier should not form [cf. \cref{eq:sigma_law_beta}]. The shaded region below the line thus corresponds to saturation via constant-flux, while above it a helicity barrier forms. The inset plot shows the time-averaged average value of the rate-of-change of the dissipation ratio \cref{eq:dissipation_ratio} normalised to $\fluxh/H$ for each set of simulations, as indicated by the colour, with the horizontal axis rescaled by the critical beta \cref{eq:sigma_law_beta}. The horizontal dotted line therein corresponds to the value \cref{eq:dissipation_ratio_critical} above which the helicity barrier is determined to have formed.}
	\label{fig:beta_scan}
\end{figure}

For the prediction (ii) --- that a helicity barrier forms only when the critical perpendicular wavenumber \cref{eq:sigma_law_kperp_crit} is well-resolved --- we consider the set of simulations labelled ``resolution scan'' in \cref{tab:simulation_parameters}. These all lie in the FLR-MHD limit (finite $\rhoi$, $d_e \rightarrow 0$). In these simulations, as well as all others considered in this paper, the adaptive dissipation (see \cref{sec:numerics}) is implemented in such a way as to ensure that the dissipation scale $\kperp^\text{diss}$ is approximately half that of the maximum wavenumber in the simulation $\kperp^\text{max}$, irrespective of forcing amplitude and injection imbalance. This means that \cref{eq:sigma_law_kperp_crit} can be re-written as a condition on $\kperp^\text{max}$, viz., 
\begin{align}
	\kperp^\text{max} \rhoi \lesssim 2 \kperp^\text{crit} \rhoi = \frac{2}{\forcing} \left(\frac{2}{1+Z/\tau}\right)^{1/2},
	\label{eq:kperp_max_condition}
\end{align}
where $\kperp^\text{crit}$ is as defined in \cref{eq:sigma_law_kperp_crit}. We expect no helicity barrier to be present if the inequality in \cref{eq:kperp_max_condition} is satisfied.
To confirm this, we varied the injection imbalance $\forcing$ across a different set of resolutions from $64^3$ to $256^3$, as in \cref{tab:simulation_parameters}. The results of this scan are shown in \cref{fig:resolution_scan}, which show good agreement with \cref{eq:kperp_max_condition} --- deviations from this prediction are due to the difficulty of ensuring adequate separation between regions of parallel and perpendicular dissipation at lower resolutions. That being said, given that real physical systems are not limited by resolution, we do not consider exact agreement with \cref{eq:kperp_max_condition} necessary. Rather, these results provide clear evidence supporting the general principles of helicity barrier formation outlined above.

\begin{figure}
	
	\includegraphics[width=1\textwidth]{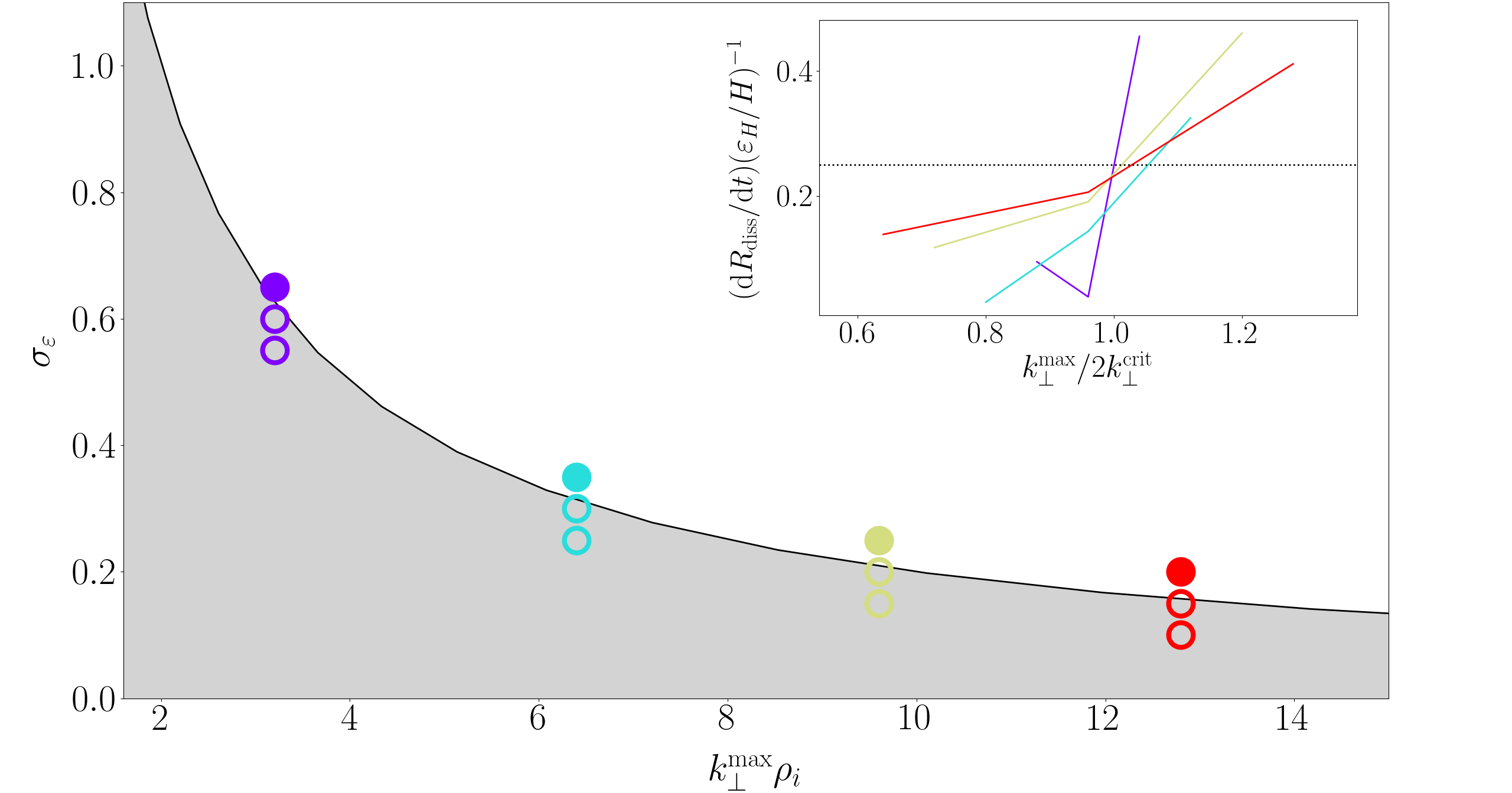}
	
	\caption[]{Data from a two-dimensional parameter scan of injection imbalance $\forcing$ versus the maximum wavenumber $\kperp^\text{max} \rhoi$ set by the numerical resolution. Each circle corresponds to a simulation from the set ``resolution scan'' in \cref{tab:simulation_parameters}, with filled/open circles indicating the presence/absence of the helicity barrier. The solid black line corresponds to the value of $\forcing$ below which, for a given $\kperp^\text{max} \rhoi$, the helicity barrier should not form [see \cref{eq:kperp_max_condition}]. As in \cref{fig:beta_scan}, the shaded region below the line thus corresponds to saturation via constant-flux, while above it a helicity barrier forms. The inset plot shows the time-averaged average value of the rate-of-change of the dissipation ratio \cref{eq:dissipation_ratio} normalised to $\fluxh/H$ for each set of simulations, as indicated by the colour, with the horizontal axis rescaled by $2\kperp^\text{crit} \rhoi$. The horizontal dotted line therein corresponds to the value \cref{eq:dissipation_ratio_critical} above which the helicity barrier is determined to have formed.}
	\label{fig:resolution_scan}
\end{figure}

\subsection{A comment on dynamic phase alignment}
\label{sec:dynamic_phase_alignment}
Before discussing the implications of these findings, let us briefly comment on the relationship between the helicity barrier and the concept of dynamic phase alignment \citep{loureiro18,milanese20}. As discussed in \cref{sec:effect_of_helicity_conservation}, dynamic phase alignment refers to the phenomenon whereby, in the ultra-low-beta regime $\beta_e \ll m_e/m_i$, the fluctuations of the electrostatic potential $\phi$ and parallel magnetic vector potential $\Apar$ become increasingly misaligned at small scales in order to maintain a constant flux of both free energy and helicity. The principal finding of \cite{milanese20} was that this alignment is directly manifest in the phase angle \cref{eq:phase_angle}, in that it becomes a decreasing function of perpendicular wavenumber, as per the theoretical prediction \cref{eq:phase_angle_estimate}. As we discussed in \cref{sec:effect_of_helicity_conservation}, the helicity barrier arises as a consequence of such an alignment becoming impossible, breaking the constant-flux solution.

 \begin{figure}
	
	\includegraphics[width=1\textwidth]{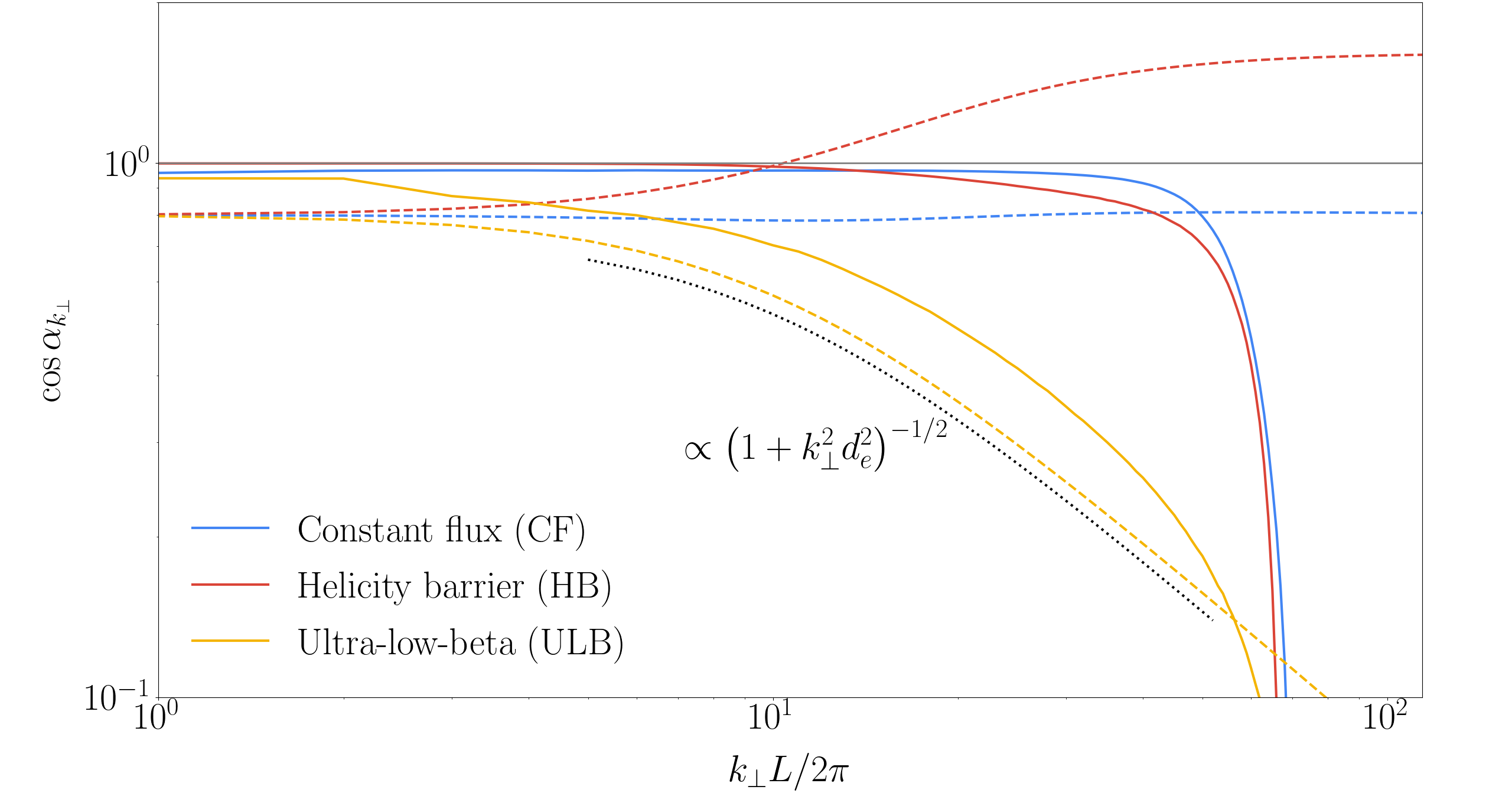}
	
	\caption[]{The phase angle \cref{eq:phase_angle} between fluctuations of the electrostatic potential $\phi$ and parallel magnetic vector potential $\Apar$, as a function of perpendicular wavenumber, for the three simulations labelled ``comparison'' in \cref{tab:simulation_parameters} (solid lines). Simulations CF and HB both have $\rhoi = 0.1 L$ and finite $d_e$, while simulation ULB has $d_e =0.1 L$ but $\rhoi \rightarrow 0$. The dashed lines show the theoretical scaling \cref{eq:phase_angle_estimate}, while the dotted black line shows the expected scaling in the ultra-low-beta regime. We do not expect exact agreement between \cref{eq:phase_angle_estimate} and \cref{eq:phase_angle} because the former is a result derived using a ratio of fluxes (see \cref{sec:effect_of_helicity_conservation}) while for the latter we are plotting a ratio of amplitudes (or, equivalently, energies).}
	\label{fig:phase_angle}
\end{figure} 

\newpage

To illustrate this, we plot, in \cref{fig:phase_angle}, the phase angle \cref{eq:phase_angle} for the set of three simulations labelled ``comparison'' in \cref{tab:simulation_parameters} --- the first two are those examined in detail in \cref{sec:the_helicity_barrier}, while the third is in the ultra-low-beta (ULB) regime considered by \cite{loureiro18} and \cite{milanese20}. It is clear that in the latter case, $\cos\phase$ decreases with perpendicular wavenumber, as demonstrated by \cite{milanese20}, while the former cases both have $\cos\phase \approx 1$: simulation CF is sufficiently close to the critical boundary predicted by \cref{eq:sigma_law} [or, alternatively, \cref{eq:phase_angle_estimate}] that it must be as highly aligned in order to maintain a constant flux, while simulation HB has already passed beyond this threshold and formed a helicity barrier, with the fluctuations remaining maximally aligned on the largest scales. We have also plotted (dashed lines) the theoretical scalings \cref{eq:phase_angle_estimate} for the phase angle in \cref{fig:phase_angle}; the fact that this curve exceeds unity for simulation HB implies the formation of a helicity barrier, i.e., it violates the criterion \cref{eq:phase_angle_estimate}, which is what we indeed find dynamically. In some sense, the helicity barrier can be viewed as the ``opposite'' of dynamic phase alignment, in that it is the state that occurs when the system cannot sufficiently align the $\phi$ and $\Apar$ fluctuations. Indeed, heuristically, these cases can easily be distinguished by recalling the definitions of the free energy \cref{eq:free_energy_thetapm} and helicity \cref{eq:helicity_thetapm}. While it always remains possible to decrease the difference between the $\thetapm$ energies at small scales to compensate for the decrease in the phase velocity at $\beta/\betacrit \lesssim 1$, the opposite is not always true at $\beta/\betacrit \gtrsim 1$: at a certain perpendicular wavenumber, the difference in energies would need to be greater than the sum, and so the constant flux solution must break down. There is, however, a crucial difference between these two cases: while dynamic phase alignment modifies the dynamics at small perpendicular scales in order to extend the constant-flux solution to lower values of $\beta_e$, the helicity barrier explicitly violates the constant-flux solution, placing it into an entirely different class of turbulent dynamics to that which can be obtained via dynamic phase alignment. Importantly, the effects of the helicity barrier are thus manifest even on the largest perpendicular scales accessible to the system, having dramatic implications for the turbulent heating properties thereof.

\section{Summary and discussion}
\label{sec:summary_and_discussion}
The findings presented in this paper serve as a detailed illustration of the sensitivity of imbalanced turbulence to small changes in its characteristic physical parameters. Starting from equations derived in the low-beta asymptotic limit of gyrokinetics (isothermal KREHM; see \cref{sec:isothermal_KREHM}), we showed that the requirement for the simultaneous conservation of both free energy and helicity implies the presence of a critical electron beta (see \cref{sec:effect_of_helicity_conservation}) given by
\begin{align}
	\betacrit = \frac{2Z}{1+\tau/Z} \frac{m_e}{m_i} \frac{1}{\forcing^2},
	\label{eq:betacrit}
\end{align} 
where $Z$ is the ion charge, $\tau$ is the equilibrium-temperature-ratio between ions and electrons, with $m_e/m_i$ their mass ratio, and $\forcing$ is the ratio of the injection rates of cross-helicity and energy at large scales (injection imbalance). This theoretical prediction is well-supported by simulations of imbalanced turbulence in isothermal KREHM across a wide range of $\beta_e$ and $\forcing$ (see \cref{fig:beta_scan}). Systems situated close to either side of \cref{eq:betacrit} in parameter space exhibit dramatically different turbulent dynamics, as evident from even the most cursory glance at the real-space turbulence snapshots shown in \cref{fig:realspace_data} (showing two simulations with values of $\beta_e$ that differ by a factor of four, but are otherwise identical). Specifically, in systems with $\beta_e$ below \cref{eq:betacrit}, the free energy injected at the largest perpendicular scales is able to undergo a constant-flux, Alfv\'enic cascade to smaller scales $\kperp \rhoe \lesssim 1$ where it dissipates, and in so doing depositing the majority of the turbulent free energy into electron heating. Systems with $\beta_e$ exceeding \cref{eq:betacrit}, on the other hand, are unable to support a constant-flux solution, inevitably forming a helicity barrier. This prevents all but the balanced portion of the injected free energy ($\approx 2 \fluxm$) from cascading past the ion Larmor scale $\kperp \rhoi \sim 1$, resulting in a build-up of turbulent free energy at larger scales $\kperp \rhoi \lesssim 1$. Fluctuations on these scales eventually form fine parallel structures which, in our system, dissipate via parallel hyperviscosity --- in a more comprehensive system, they would excite high-frequency ion-cyclotron waves (ICWs), leading to perpendicular ion heating \citep{squire22,squire23}. Thus, the constant-flux and helicity-barrier states, demarcated by the critical beta \cref{eq:betacrit}, offer entirely different propositions for turbulent heating: the majority of the injected turbulent free energy is converted into electron heating, in the former case, or ion heating, in the latter.

Assuming that the existence of the helicity barrier, or otherwise, plays a central role in determining turbulent heating, these results have clear implications for observations of imbalanced Alfv\'enic turbulence. For highly imbalanced plasmas with a modest plasma beta, we would expect to observe dominant ion heating, and for the spectral slopes of the electromagnetic fields to exhibit a steep ``transition range'' scaling $\sim \kperp^{-4}$ bracketing $\kperp \rhoi \sim 1$, a distinctive feature of helicity-barrier-mediated turbulence \citep{meyrand21,squire22,squire23}. Conversely, for plasmas with either a small imbalance (at large solar radii) or a very low plasma beta ($\beta_e$ approaching $m_e/m_i$), we would expect to observe more electron heating, and for the steep transition range scaling to be replaced by the much shallower $\sim \kperp^{-7/3}$ scaling predicted from KAW turbulence (see \cref{sec:constant_flux_cascade}). Given that much of the solar wind typically has $m_e/m_i \ll \beta_e \lesssim 1$ \citep{bruno05}, \cref{eq:betacrit} would suggest that it should usually display signatures of helicity-barrier-mediated turbulence. This is consistent with observations: ions are typically hotter than electrons \citep{cranmer09}, with significant power in ICWs around $\kpar \rhoi \sim 1$  \citep{huang20,bowen20waves}, while spectra observed by PSP usually show the aforementioned steep transition range scalings \citep{bowen20,duan21,bowen24}. 

Nevertheless, there may be regions of the solar wind in which the helicity barrier does not operate. Equation \cref{eq:betacrit} predicts that a constant-flux cascade is always possible if $\beta_e \leqslant m_e/m_i$, irrespective of imbalance, as well as the fact that the helicity barrier will not form at sufficiently low imbalance. This means that the helicity barrier is unlikely to operate in regions of the solar wind that are strongly magnetised or have low imbalance, or indeed some combination of the two. That being said, the values of $\forcing$ below which the helicity barrier will not operate are quite low. Indeed, our simulations showed helicity-barrier formation at values of the injection imbalance as low as $\forcing = 0.2$ (see \cref{fig:beta_scan}), with lower values limited by numerical resolution, rather than by some fundamental constraint on the dynamics (see \cref{sec:effect_of_helicity_conservation} or the last paragraph in \cref{sec:breaking_the_helicity_barrier}). It is worth noting in this context, however, that the injection imbalance $\forcing$ is not the same as the often-measured normalised cross-helicity $\sigma_c$, so careful work is required in order to extract an exact correspondence between values of $\forcing$ used in simulations and the dynamics observed in astrophysical systems.
}

\subsection{Limitations of the isothermal approximation}
\label{sec:limitations_of_isothermal}
An important limitation of this current study concerns the possible role of electron kinetics, which we have here neglected in order to isolate the effects of including finite electron inertia. The inclusion thereof has two important physical consequences that could potentially alter the results above: (i) it introduces electron Landau damping as another channel for turbulent electron heating, allowing energy to be transferred to small scales in velocity space (in addition to in wavenumber space); (ii) it modifies the conservation of helicity viz., instead of \cref{eq:helicity} being everywhere conserved, it is able to be injected and/or removed by higher-order velocity moments of the kinetic distribution function. Both of these effects are most significant around $\kperp d_e \sim 1$, and so we expect little change for values of $\beta_e$ significantly above \cref{eq:betacrit}. For those close to \cref{eq:betacrit}, however, the dynamics could be significantly modified, e.g., the helicity barrier may not form because the helicity ceases to be globally conserved or, should the helicity barrier persist, electron Landau damping could provide another source of dissipation for fluctuations at $\kperp \rhoi \lesssim 1$. This latter effect would be significant, as the amplitude reached by fluctuations on these scales plays a central role in determining the amount of perpendicular ion heating by ICWs \citep{squire22,squire23}, and thus the effective fraction of the injected energy earmarked for electron heating. In either case, the re-introduction of electron kinetics should have the effect of shifting the critical beta \cref{eq:betacrit} towards larger values (i.e., moving the solid black curve in \cref{fig:beta_scan} to the right), giving rise to more electron, and less ion, heating. Confirming these predictions is the subject of ongoing work. In perhaps a preview of what is to come, recent simulations of balanced turbulence by \cite{zhou23PNAS} suggest that the advection of energy in velocity space is the dominant mechanism of nonlinear energy transfer at $\kperp d_e \sim 1$, which becomes the primary route to dissipation. 

There are, of course, other mechanisms for turbulent heating that may be playing a role. A compressive energy cascade, although unable to exchange energy with Alfv\'enic motions \citep{sch19}, is able to cause parallel ion heating \citep{sch09}, and will break helicity conservation around $\kperp \rhoi \sim 1$, potentially arresting the breakdown of the constant flux solution associated with helicity-barrier formation. In order for either of these effects to be significant, however, the powers in the compressive and Alfv\'enic cascades would likely need to be comparable, which is generally not observed in the solar wind \citep{bruno13,chen16,chen20}. Other heating mechanisms that rely on the presence of large-amplitude fluctuations, such as stochastic heating \citep{chandran10} or sub-ion-Larmor-radius KAW turbulence \citep{arzamasskiy19,isenberg19}, may also play a role.

Even with these considerations taken into account, the results of this paper represent a robust prediction of two fundamentally different types of turbulence: one of a constant-flux cascade of energy to small scales, the other involving the build-up and dissipation of energy at the largest scales due to the helicity-barrier mechanism. The critical beta \cref{eq:betacrit} thus marks a boundary between two dramatically different regimes of turbulent heating, each with clear observational signatures that can further constrain the possible physical processes at play in imbalanced Alfv\'enic turbulence. This physics could play an important role in many magnetised astrophysical environments where the same symmetry exists (i.e., having imbalance), such as accretion disks, the intracluster medium, and the solar wind context discussed in detail in this work. Therefore, advancing our understanding of the dynamics related to the helicity barrier and the resultant mechanisms of turbulent heating will carry important implications for plasma turbulence across a diverse array of astrophysical contexts.


\section*{Acknowledgements}
The authors would like to thank B.~D.~G. Chandran, Z. Johnston, M.~W.~Kunz, and A.~A. Schekochihin for helpful discussions and suggestions at various stages of this project. 

\section*{Funding}
The authors acknowledge the support of the Royal Society Te Ap\=arangi, through Marsden-Fund grant MFP-UOO2221 (TA and JS) and MFP-U0020 (RM), as well as through the Rutherford Discovery Fellowship RDF-U001004 (JS). Computational support was provided by the New Zealand eScience Infrastructure (NeSI) high-performance computing facilities, funded jointly by NeSI's collaborator institutions and through the Ministry of Business, Innovation \& Employment's Research Infrastructure programme.

\section*{Declaration of interests}
The authors report no conflicts of interest.


%
%
%
%

\bibliography{bibliography.bib}
\bibliographystyle{jpp}         
	
\end{document}